\documentclass[letterpaper, 10 pt, conference]{ieeeconf}  % Comment this line out
\IEEEoverridecommandlockouts                              % This command is only
\usepackage{amsmath}
\usepackage{color}
\usepackage{subfigure}
\usepackage{graphicx,epsfig,amsfonts,bbm,epstopdf,tabularx}
\usepackage{ifthen}
\usepackage{cite}

\newcommand{\be}{\begin{equation}}
\newcommand{\ee}{\end{equation}}
\newcommand{\bee}{\begin{eqnarray}}
\newcommand{\eee}{\end{eqnarray}}
\newcommand{\bse}{\begin{subequations}}
\newcommand{\ese}{\end{subequations}}
\newcommand{\ins}{\boldsymbol{\omega}}

\newtheorem{lemma}{\textbf{Lemma}}
\newtheorem{theorem}{\textbf{Theorem}}

\newtheorem{definition}{\textbf{Definition}}

\ifodd 1
\newcommand{\com}[1]{\textbf{\color{blue} (COMMENT: #1)}}%comment of the text
\else

\newcommand{\com}[1]{}
\fi

\definecolor{darkgreen}{rgb}{0,0.5,0}
\newboolean{showcomments}
\setboolean{showcomments}{true}
\newcommand{\desmond}[1]{\ifthenelse{\boolean{showcomments}}
{ \textcolor{red}{(Desmond says:  #1)}}{}}

\newcommand{\hide}[1]{}

\title{\Large \bf
Understanding the Inefficiency of Security-Constrained Economic Dispatch
}

\author{Mohammad H. Hajiesmaili$^{1}$, Desmond Cai$^{2}$, and Enrique Mallada$^{1}$% <-this % stops a space
	\thanks{$^{1}$M. Hajiesmaili and E. Mallada are with the Department of Electrical and Computer Engineering, the Johns Hopkins University, emails:
		{\tt\small \{hajiesmaili,mallada\}@jhu.edu}}%
	\thanks{$^{2}$D. Cai is with the Institute of High Performance Computing, A*STAR, Singapore, email: {\tt\small desmond-cai@ihpc.a-star.edu.sg}}%
}

\begin{document}

\maketitle
\thispagestyle{empty}
\pagestyle{empty}

%%%%%%%%%%%%%%%%%%%%%%%%%%%%%%%%%%%%%%%%%%%%%%%%%%%%%%%%%%%%%%%%%%%%%%%%%%%%%%%%
\begin{abstract}
The security-constrained economic dispatch (SCED) problem tries to maintain the reliability of a power network by ensuring that a single failure does not lead to a global outage. The previous research has mainly investigated SCED by formulating the problem in different modalities, e.g. preventive or corrective, and devising efficient solutions for SCED. In this paper, we tackle a novel and important direction, and  analyze the economic cost of incorporating security constraints in economic dispatch. Inspired by existing inefficiency metrics in game theory and computer science, we introduce notion of \textit{price of security} as a metric that formally characterizes the economic inefficiency of security-constrained economic dispatch as compared to the original problem without security constraints. Then, we  focus on the preventive approach in a simple topology comprising two buses and two lines, and investigate the impact of generation availability and demand distribution on the price of security. Moreover, we explicitly derive the worst-case input instance that leads to the maximum price of security. By extensive experimental study on two test-cases, we verify the analytical results and provide insights for characterizing the price of security in general networks. 

\end{abstract}

\section{INTRODUCTION}

The primary goals in power system operation are to minimize operating costs and maintain system reliability~\cite{wood2012power}. The economic dispatch (ED) problem minimizes generation costs subject to operating constraints~\cite{Zhang2016Peak,cai2016distributed}. To ensure that failures do not cascade after major disturbances, such as line or generator outages, system operators add security constraints to the economic dispatch problem~\cite{kundur2004definition}. The resulting problem is known as security-constrained economic dispatch (SCED)~\cite{alsac1974optimal,aoki1982economic,stott1987security,somasundaram2004evolutionary,capitanescu2007contingency}. The typical criteria is that the system must be robust to the failure of any single element, i.e. the solution must satisfy the $N-1$ condition~\cite{ejebe1979automatic}. 

There are currently two major approaches to SCED. Preventive approaches impose additional operating limits for the post-disturbance configurations, resulting from contingencies, without taking into account the corrective capabilities of the system~\cite{capitanescu2007contingency}. In contrast, corrective approaches leverage the system's real-time corrective capabilities after an outage, such as generation rescheduling, switching, congestion management, etc.~\cite{Monticelli1987Security}. Therefore, while preventive approaches are simpler to implement than corrective approaches, the former are overly conservative and more expensive. Nevertheless, majority of SCED implementations today are preventive. Historically, this may be due, in part, to more complex control, sensing, and communication requirements of real-time corrective dispatch. However, recent research has demonstrated that it is possible to efficiently dispatch generators in real-time and distributed manners to rapidly correct for grid disturbances~\cite{ZhaoMalladaLowBialek2016,2014arXiv1410.2931M,cai2016distributed}. 

With the growth of renewables and distributed generation, existing approaches for ensuring system security may not be appropriate for the future grid. The inefficiency of preventive approaches could become more significant due to increased operating uncertainty and greater number of active generation sources. The future grid is also more likely to have multiple correlated failures, which necessitates additional contingency considerations beyond $N-1$~\cite{kaplunovich2013fast}, leading to even more conservative scheduling and higher costs. Therefore, it is increasingly important to understand the tradeoffs between different approaches for ensuring security. Specifically, there is a need to understand the impact of security constraints on operating costs and their tradeoffs against the benefits of system reliability. 

To date, we are not aware of any analysis of the operating costs attributable to security constraints. While there is a large body of literature on SCED, majority of the research have focused on developing efficient algorithms for solving the problem~\cite{ejebe1979automatic,aoki1982economic,somasundaram2004evolutionary,capitanescu2007contingency,wang2013computational,liu2015computational}. This is motivated by the fact that the size of the problem increases significantly when security constraints are added to the economic dispatch problem. Understanding the additional costs incurred due to security constraints, as well as how the costs depend on system structure (e.g. network topology, demand profiles, generation availability, etc.), may also provide insights into the most critical components in the system, and in turn guide resource allocation, maintenance decisions, and infrastructure investments.

In this paper, we study the impact of security constraints on operating costs. We focus on preventive approaches as it is the most prevalent approach for ensuring security in current power systems. In particular, we study the cost of ensuring $N-1$ security by investigating the ratio of dispatch costs at the solution of SCED to that at the solution of ED (i.e. removing security constraints from SCED). We refer to this ratio as the price of security. We completely characterize the price of security for a simple topology comprising two buses and two lines. 

Our analyses illustrates a few phenomena. First, the price of security always increases when there is more cheap generation capacity in the system. 
%This could be expected since more usage of cheap generation implies that cheap generation is substituted for expensive generation in order to ensure security. 
Second, the price of security is maximized when the lines between the two buses are saturated. 
This could be expected since the most cheap generation is substituted for expensive generation (to ensure security) when the lines are most heavily utilized. However, our analyses also reveals a counter-intuitive phenomenon. Given fixed total demand, having more demand distributed on the cheaper node may in fact increase the price of security. This occurs when the transmission line is fully utilized, and so additional demand does not change the cost of ensuring security; but when more demand is distributed on the cheaper node, the optimum economic dispatch cost is smaller, and therefore the cost of security has a relatively bigger impact on the dispatch cost. 

Finally, we investigate numerically the price of security for the PJM $5$-bus system~\cite{Li2010Small} and illustrate that some of our theoretical results in the $2$-bus case manifest in more general settings. In particular, the numerical results on the PJM $5$-bus system show that the impact of generation capacity and demand distribution at the cheap region of the network is similar to that of the $2$-bus system. However, finding the worst-case demand that maximizes the price of security is a formidable task that depends on properties of the lines between two regions, aggregate demand, and demand distribution. 
% \desmond{Any more specific examples? Or new insights not present in the original 2-bus case?}

\section{System Model}
In this section, we introduce the power system model used throughout this paper and define the economic dispatch and security-constrained economic dispatch problems. Then we define the proposed metric for measuring the inefficiency of security-constrained economic dispatch.

\subsection{Power System}

We model the topology of the power network by a directed graph\footnote{Note that in reality the power flow on the links are bidirectional, however, it is a common practice to model the network topology as a directed graph with arbitrary directions on the edges.} $\mathcal{G} = (\mathcal{V},\mathcal{E})$, where $\mathcal{V}$ is the set of nodes (buses, used interchangeably), indexed by $v$, and $\mathcal{E}$ is the set of edges (lines or branches, used interchangeably), indexed by $e$. Let $n := |\mathcal{V}|$ and $m:=|\mathcal{E}|$ denote the number of nodes and edges respectively.

We assume that each node $v$ has exactly one generator and one load.\footnote{Considering linear cost model for generators, this assumption is not restrictive multiple generators (resp. loads) at a node can be equivalently represented by a single generator (resp. load) via an appropriate transformation of costs (resp. demands).} Assume that the generator at node $v$ has a maximum generation capacity $\overline{q}_v\in\mathbf{R}_+$, and it incurs a cost $\alpha_v q_v$ when generating $q_v$, where the coefficient $\alpha_v \in\mathbf{R}_+$. Let $d_v$ denote the demand at node $v$. Define the vectors $\overline{\mathbf{q}} := (\overline{q}_v,v\in\mathcal{V})$, $\mathbf{q} := (q_v,v\in\mathcal{V})$, and $\boldsymbol\alpha := (\alpha_v,v\in\mathcal{V})$. 

Let $f_e$ denote the power flow on edge $e$ and assume that the edge has a thermal line limit (capacity) $\overline{f}_e$. Define the vectors ${\overline{\mathbf{f}} := (\overline{f}_e,e\in\mathcal{E})}$ and ${\mathbf{f} := (f_e, e\in\mathcal{E})}$. We assume a DC power flow model and let $\mathbf{H}$ be the $m\times n$ matrix of shift factors that map power injections to line flows. Then the latter are given by
\begin{eqnarray}
\mathbf{f} = \mathbf{H} (\mathbf{q}-\mathbf{d}).
\end{eqnarray}

\subsection{Economic Dispatch}
The economic dispatch problem minimizes generation costs subject to operating constraints and is given by:
\bse
\bee
\textsf{ED}:  \min_{\mathbf{q}}  && \boldsymbol\alpha^{\mathsf{T}}\mathbf{q} \\
\text{s.t.} && \mathbf{0} \leq \mathbf{q} \leq \overline{\mathbf{q}},\label{eq:gen_cap} \\
&& \boldsymbol{1}^{\mathsf{T}}(\mathbf{q} - \mathbf{d}) = 0,\label{eq:sup_dem_bal} \\
&&  -\overline{\mathbf{f}} \leq  \mathbf{H} (\mathbf{q}-\mathbf{d}) \leq \overline{\mathbf{f}}. \label{eq:line_cap} 
%\text{vars.} && \mathbf{q},\nonumber
\eee
\ese
Constraint~\eqref{eq:gen_cap} restricts generations to capacities, constraint~\eqref{eq:sup_dem_bal} enforces supply-demand balance, and constraint~\eqref{eq:line_cap} restricts line flows to line limits. 

\subsection{Security-Constrained Economic Dispatch}
Next, we formulate the security-constrained economic dispatch problem. In this paper, we focus on robustness to the outage of any single line. Associate with the outage of an edge $e \in \mathcal{E}$ the $m-1$ vector $\overline{\mathbf{f}}_{-e} = (f_{e'}:e'\in\mathcal{E}, e'\neq e)$ of line capacities and the $(m-1)\times n$ matrix $\mathbf{H}_{-e}$ of shift factors. We are interested in the following security-constrained economic dispatch problem: 
\bse
\bee
\textsf{SCED}:  \min_{\mathbf{q}}  &&  \boldsymbol\alpha^\mathsf{T}\mathbf{q}\\
\text{s.t.} && \mathbf{0} \leq \mathbf{q} \leq \overline{\mathbf{q}},\label{eq:gen_cap2} \\
&& \boldsymbol{1}^{\mathsf{T}}(\mathbf{q} - \mathbf{d}) = 0,\label{eq:sup_dem_bal2} \\
&&  -\overline{\mathbf{f}} \leq  \mathbf{H} (\mathbf{q}-\mathbf{d}) \leq \overline{\mathbf{f}}, \label{eq:line_cap2}\\ 
&&  -\overline{\mathbf{f}}_{-e} \leq  \mathbf{H}_{-e} (\mathbf{q}-\mathbf{d}) \leq \overline{\mathbf{f}}_{-e}, \forall e. \label{eq:line_cap_outage}
%\text{vars.} && \mathbf{q},\nonumber
\eee
\ese
Note that \textsf{SCED} contains $2m(m-1)$ more constraints than \textsf{ED}, which are represented by~\eqref{eq:line_cap_outage}, each of which is associated with a unique line outage. 

\subsection{Price of Security}
\label{sec:pos}
Our goal is to understand the cost of ensuring security to outage of any single line. To that end, we define a metric to compare the operating costs of the solutions to \textsf{ED} and \textsf{SCED}. Our metric has an intuitive interpretation and is motivated by inefficiency metrics in game theory and computer science (e.g. price of anarchy, price of stability, competitive ratio)~\cite{nisan2007algorithmic,borodin2005online}. 

Given a network $\mathcal{G}$, cost coefficients $(\alpha_v, v\in\mathcal{V})$, and transmission line limits $(f_e, e\in\mathcal{E})$, let ${\ins =  (\mathbf{\overline{q}}, \mathbf{d}) \in \Omega}$ be an input instance to \textsf{ED} and \textsf{SCED}, where $\Omega$ is the set of all possible different instances of generation capacities and demands that are feasible to both problems. 
%We only consider input instances $\ins$ that are feasible for both \textsf{ED} and \textsf{SCED}. 
Define $c^{\star}_{\textsf{ed}}(\ins)$ and $c^{\star}_{\textsf{sc}}(\ins)$ as the optimal values of \textsf{ED} and \textsf{SCED} under input instance $\ins$, respectively. 
\begin{definition}
	We define the \emph{price of security} for input instance $\ins$ by: 
	\begin{equation}
	\label{eq:pos_def}
	\textsf{PoS}(\ins) := \frac{c^{\star}_{\textsf{sc}}(\ins)}{c^{\star}_{\textsf{ed}}(\ins)}.
	\end{equation}
\end{definition}

Note that  all feasible solutions of \textsf{SCED} are also feasible for \textsf{ED}, hence, it follows that $c^{\star}_{\textsf{ed}}(\ins) \leq c^{\star}_{\textsf{sc}}(\ins)$, and hence, ${\textsf{PoS}(\ins) \geq 1}$. We are interested in characterizing the instance that lead to the largest value for $\textsf{PoS}(\ins)$, that is, the maximum extra cost of ensuring security. 

Since it is typically difficult to obtain closed form expressions for the solutions to \textsf{ED} and \textsf{SCED} (as a function of $\omega$), obtaining a closed form expression for $\textsf{PoS}(\ins)$ is a challenging task in general. Moreover, system operators typically do not have direct control on demand, and generation availability varied over time. Thus, it is of interest to characterize the worst-case  generation availability and demand profile that maximizes the price of security.
\begin{definition}
	We define the worst-case price of security over all instances in $\Omega$ by:
	\begin{equation}
	\label{eq:pos_global}
	\textsf{PoS} := \sup_{\ins \in\Omega} \textsf{PoS}(\ins).
	\end{equation}
\end{definition}
%Our focus in this paper is to study the impact of network topology, cost coefficients, and transmission line limits, on the worst-case price of security. 

Note that we define the worst-case price of security over different generation capacities and demands only, assuming fixed network topology, cost functions, and line limits. This is motivated by the fact that the latter are typically constant over longer time-scales (i.e. days or months) while generation availabilities and demands vary greatly over shorter time-scales (i.e. hours). Moreover, as we will demonstrate in this paper, generation capacities and demands alone have complicated and surprising impacts on the price of security. Therefore, we focus on analyzing the worst-case price of security over generation capacities and demands, and leave the analyses with respect to other factors to future work. 

To obtain insights into the problem, we begin by analyzing a simple 2-bus topology in the next section. Our analyses provide insights into the major determinants of the costs of security. Then, in the subsequent section, we investigate these insights numerically on the 2-bus topology as well as the PJM 5-bus system~\cite{Li2010Small}

\section{Analysis of 2-Bus Topology \label{sec:2bus}}
\begin{figure}[!t]
	\centering
	\includegraphics[scale=0.3]{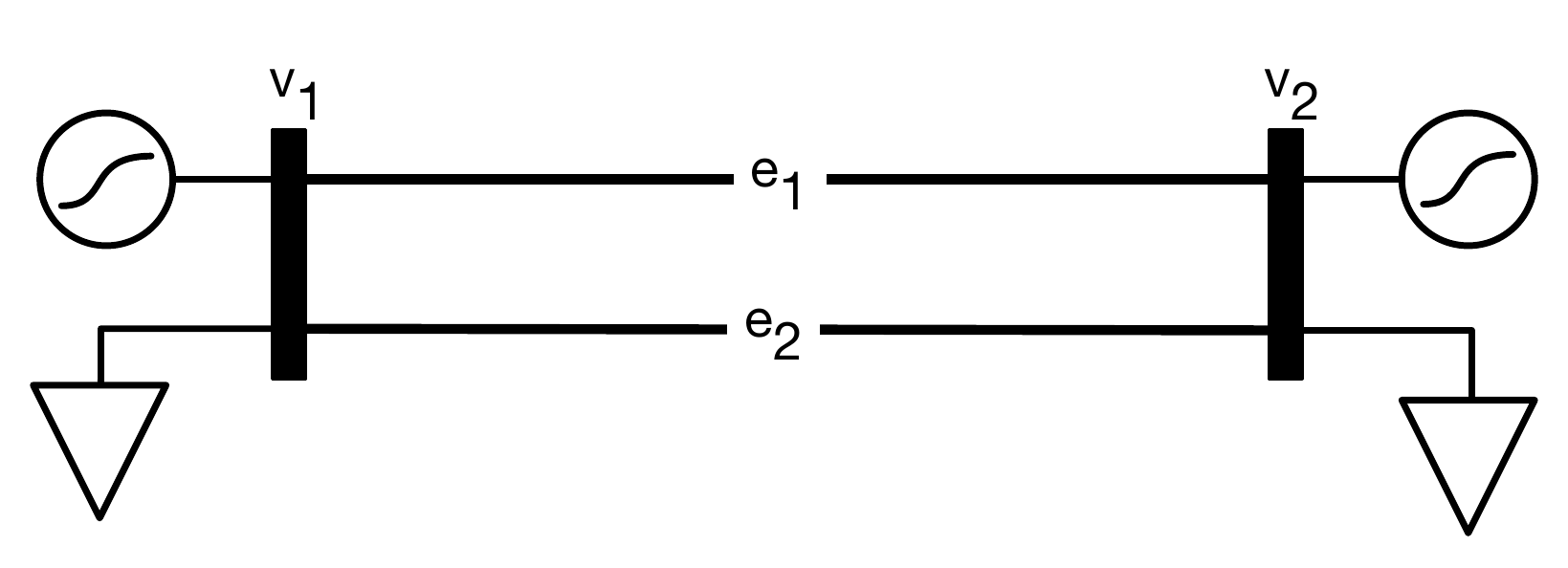}
	\vspace{-5mm}\caption{A simple $2$-bus topology}
	\label{fig:2bus}
\end{figure}

In this section, we analyze the price of security of the simple $2$-bus topology shown in Fig.~\ref{fig:2bus}, where there are $2$ nodes connected by $2$ edges. Therefore, $\mathcal{V} = \{v_1,v_2\}$ and $\mathcal{E} = \{e_1,e_2\}$. Our results provide insights into behavior in more general topologies.
All proofs are given in Appendix.
%The result for this simple topology could be considered as the road-map for the more general network topologies. 

First, we simplify \textsf{ED}. By specializing \textsf{ED} to the $2$-bus topology, we obtain the following problem:
\bse
\bee
&&\textsf{ED-2b}:\nonumber\\
\min_{q_1,q_2}  &&   \alpha_1q_1 + \alpha_2q_2 \nonumber\\
\text{s.t.} && 0 \leq q_1 \leq \overline{q}_1,\label{eq:gen_cap_2bus} \\
&& 0 \leq q_2 \leq \overline{q}_2,\label{eq:gen_cap_2bus_2} \\
&& (q_1 - d_1) + (q_2 - d_2)  = 0,\label{eq:sup_dem_bal_2bus} \\
&&  -\overline{f}_1 \leq  \frac{B_1(q_1-d_1 - q_2 +d_2)}{2(B_1+B_2)}  \leq \overline{f}_1, \label{eq:line_cap_2bus} \\
&&  -\overline{f}_2 \leq  \frac{B_2 (q_1-d_1 - q_2 +d_2)}{2(B_1+B_2)}  \leq \overline{f}_2, \label{eq:line_cap_2bus_2} 
%\text{vars.} && \mathbf{q},\nonumber
\eee
\ese
where constraints~\eqref{eq:gen_cap_2bus} and~\eqref{eq:gen_cap_2bus_2} are equivalent to the generation capacity constraint~\eqref{eq:gen_cap} in \textsf{ED}, constraint~\eqref{eq:sup_dem_bal_2bus} is the supply-demand balance constraint, and constraints~\eqref{eq:line_cap_2bus} and~\eqref{eq:line_cap_2bus_2} are the line constraints associated with lines $e_1$ and $e_2$, respectively. By substituting equation~\eqref{eq:sup_dem_bal_2bus} into inequalities~\eqref{eq:line_cap_2bus} and~\eqref{eq:line_cap_2bus_2}, the latter two inequalities are equivalent to the following single constraint:
\begin{equation}
\label{eq:fhat}
-f^{\textsf{ed}} \leq q_1 - d_1 \leq f^{\textsf{ed}},
\end{equation}
where 
\begin{equation}
\label{eq:fed}
f^{\textsf{ed}} := (B_1+B_2)\min \left\{\frac{\overline{f}_1}{B_1},\frac{\overline{f}_2}{B_2}\right\}.
\end{equation}
Note that $f^{\textsf{ed}}$ can be interpreted as the maximum flow from node $1$ to node $2$.

Next, we simplify \textsf{SCED}. By specializing \textsf{SCED} to the 2-bus topology, and making use of the simplification in~\eqref{eq:fhat}, we obtain the following problem: 
\bse
\bee
&&\textsf{SCED-2b}:\nonumber\\
\min_{q_1,q_2}  &&   \alpha_1q_1 + \alpha_2q_2 \nonumber\\
\text{s.t.} && 0 \leq q_1 \leq \overline{q}_1,\label{eq:gen_cap_2bus_sc} \\
&& 0 \leq q_2 \leq \overline{q}_2,\label{eq:gen_cap_2bus_2+sc} \\
&& (q_1 - d_1) + (q_2 - d_2)  = 0,\label{eq:sup_dem_bal_2bus_sc} \\
&&  -f^{\textsf{ed}} \leq q_1 - d_1 \leq f^{\textsf{ed}}, \label{eq:fhat_sc} \\
&&  -\overline{f}_1 \leq  1/2(q_1-d_1 - q_2 +d_2)  \leq \overline{f}_1, \label{eq:line_cap_2bus_sc_1}\\
&&  -\overline{f}_2 \leq  1/2(q_1-d_1 - q_2 +d_2)  \leq \overline{f}_2. \label{eq:line_cap_2bus_2_sc_2} 
%\text{vars.} && \mathbf{q},\nonumber
\eee
\ese
Note that \textsf{SCED-2b} contains four more constraints than \textsf{ED-2b} -- \eqref{eq:line_cap_2bus_sc_1} and~\eqref{eq:line_cap_2bus_2_sc_2} -- that reflect the outage of lines $e_2$ and $e_1$ respectively. By using a procedure similar to that which we used to derive~\eqref{eq:fhat}, we can rewrite~\eqref{eq:line_cap_2bus_sc_1} and~\eqref{eq:line_cap_2bus_2_sc_2} into the following compact form:
\begin{eqnarray}
\label{eq:f}
-f^{\textsf{sc}} \leq  q_1-d_1  \leq f^{\textsf{sc}},
\end{eqnarray}
where 
\begin{equation}
\label{eq:fsc}
f^{\textsf{sc}} := \min \{\overline{f}_1,\overline{f}_2\},
\end{equation}
is the maximum flow from node $1$ to node $2$. 

We now proceed to analyze the price of security. Recall that this is defined as the largest ratio between the optimal values of \textsf{SCED-2b} and \textsf{ED-2b}. Without loss of generality, we assume for the rest of this section that $\alpha_1 \leq \alpha_2$, i.e. the generation cost at node $1$ is cheaper than that at node $2$. We also refer to the generator at node $1$ as the cheap generator and the generator at node $2$ as the expensive generator. %Furthermore, we use $d := d_1 + d_2$ to denote the total demand in the system.

\subsection{Impact of Generation Capacities \label{sec:cap_imp}}

The following lemma highlights the impact of cheap generation availability on the price of security.
\begin{lemma}
	\label{lem:gen}
	Let $\ins = (\mathbf{\overline{q}}, \mathbf{d})$ and $\ins' = (\mathbf{\overline{q}}', \mathbf{d})$ be two input instances with identical demand profiles $\mathbf{d}$. If $\overline{q}'_1 \leq \overline{q}_1$, then $\textsf{PoS}(\ins') \leq \textsf{PoS}(\ins)$.
\end{lemma}

To investigate Lemma~\ref{lem:gen}, let us consider the case where ${\overline{q}'_1 \leq d_1+d_2  \leq \overline{q}_1}$, which implies that the total demand $d_1+d_2$ can be fully served by the cheap generator in instance $\ins$, but cannot be fully served by the cheap generator in instance $\ins'$. Lemma~\ref{lem:gen} implies that, keeping all other factors constant, the price of security is greater when the cheap generation is not limited (i.e. $\overline{q}_1 \geq d_1 + d_2$) versus when cheap generation is limited (i.e. $\overline{q}_1' \leq d_1 + d_2$). Therefore, the price of security is higher when there is greater availability of cheap generation. This is, perhaps, expected since more cheap generation is substituted for expensive generation in order to ensure security. 

Since we are interested in identifying the instances with the worst-case price of security, for the rest of our analyses, we focus on cases in which the capacity of cheap generation is greater than or equal to the total demand. 

\subsection{Impact of Demands \label{sec:dem_imp}}

Next, we focus on the impact of the demand profile on the price of security. Let $\ins = (\overline{\mathbf{q}},\mathbf{d})$ be an instance such that ${\overline{q}_1 \geq d_1 + d_2}$, i.e. all demand can be served by cheap generation. We proceed to calculate the optimal solutions of \textsf{ED-2b} and \textsf{EDSC-2b} as well as closed-form expressions for $c^{\star}_{\textsf{ed}}(\ins)$ and $c^{\star}_{\textsf{sc}}(\ins)$.

First, we compute $c^{\star}_{\textsf{ed}}(\ins)$. Recall that $f^{\textsf{ed}}$ defined in~\eqref{eq:fed} can be interpreted as the maximum flow from node $1$ (with cheap generation) to node $2$ (with expensive generation). Note that there is sufficient cheap generation to serve all demand. Therefore, the optimal solution of \textsf{ED-2b} is to serve the demand $d_1$ at node $1$ locally using cheap generation, use as much of the cheap generation as possible to serve the demand $d_2$ at node $2$, i.e. $\min\{d_2, f^{\textsf{ed}}\}$, and serve the remaining demand at node $2$ locally using expensive generation, i.e. $[d_2-f^{\textsf{ed}}]^+$, where $[\cdot]^+$ denotes the projection onto the nonnegative orthant. It follows that the optimal cost of the economic dispatch problem is given by: 
\begin{equation}
c^{\star}_{\textsf{ed}}(\ins) = \alpha_1 (d_1+\min\{f^{\textsf{ed}}, d_2\}) + \alpha_2 [d_2-f^{\textsf{ed}}]^+.\label{eq:cost_ed}\\
\end{equation}

Next, we compute $c^{\star}_{\textsf{sc}}(\ins)$. Similarly, recall that $f^{\textsf{sc}}$ defined in~\eqref{eq:fsc} can be interpreted as the  maximum flow from node $1$ to node $2$. Therefore, the optimal cost of the security-constrained economic dispatch problem is given by:
\begin{equation}
c^{\star}_{\textsf{sc}}(\ins) = \alpha_1 (d_1+\min\{f^{\textsf{sc}}, d_2\}) + \alpha_2 \label{eq:cost_sced} [d_2-f^{\textsf{sc}}]^+.
\end{equation}

It follows that the price of security for instance $\ins$ defined in~\eqref{eq:pos_def} is given by:
\begin{eqnarray}
\label{eq:pos2bus}
\textsf{PoS}(\ins) = \frac{\alpha_1 (d_1+\min\{f^{\textsf{sc}}, d_2\}) + \alpha_2 [d_2-f^{\textsf{sc}}]^+}{\alpha_1 (d_1+\min\{f^{\textsf{ed}}, d_2\}) + \alpha_2 [d_2-f^{\textsf{ed}}]^+}.
\end{eqnarray}

Observe that $\textsf{PoS}(\ins)$ is small in both low and high load regimes. This is intuitive. In the low load regime, i.e. when $d_1 + d_2 \ll f^\textsf{ed}$, the line limits are not saturated. Therefore, security to outages of any single line is unlikely to increase the dispatch cost significantly. From the definitions in~\eqref{eq:fed} and~\eqref{eq:fsc}, note that $f^\textsf{ed} \leq 2f^\textsf{sc}$. In the high load regime, i.e. $d_1 + d_2 \gg f^\textsf{ed}$, the expensive generator contributes substantially towards satisfying demand even in \textsf{ED}. Hence, security to outages of any single line has a small impact on the dispatch cost, since in both~\eqref{eq:cost_ed} and~\eqref{eq:cost_sced}, the second terms are dominant. 

The next lemma highlights the impact of cheap demand on the price of security.
\begin{lemma}
	\label{lem:dem2}
	Let ${\ins = (\mathbf{\overline{q}}, \mathbf{d})}$ and ${\ins' = (\mathbf{\overline{q}}, \mathbf{d}')}$ be two input instances such that $\overline{q}_1\geq d_1 + d_2$ and $d_2=d'_2$. 
	If  $d_1' \geq d_1$, then $\textsf{PoS}(\ins') \leq \textsf{PoS}(\ins)$. 
\end{lemma}

Lemma~\ref{lem:dem2} implies that, given a fixed demand at the expensive node, the price of security is greatest when there is no demand at the cheap node. This is, perhaps, expected since there is no additional cost to ensure security when demand is being served locally (which is the case with demand located at the cheap node). However, Lemma~\ref{lem:dem2}  does not specify which distributions of demand (over the two nodes) lead to the greatest price of security. We characterize the latter in the following lemma.
\begin{lemma}
	\label{lem:dem}
	Fix the total demand $d$ and assume that ${\overline{q}_1 \geq d_1 + d_2 = d}$. Then, the demand distribution ${\mathbf{d}=(d_1=d-d_2,d_2 = \min\{d,f^{\textsf{ed}}\})}$ yields the maximum price of security, whose value is given by:
	\begin{equation}
	\label{eq:pos_opt}
	\textsf{PoS}(\ins) = \frac{\alpha_1 (d_1+\min\{f^{\textsf{sc}}, d_2\}) + \alpha_2 [d_2-f^{\textsf{sc}}]^+}{\alpha_1 d}.
	\end{equation}
\end{lemma}

Lemma~\ref{lem:dem} states that, given fixed total demand, the price of security is largest when demand is distributed to the expensive node, but only until the total demand is up to $f^{\textsf{ed}}$. When total demand increases beyond $f^{\textsf{ed}}$, having more demand distributed on the cheap node can, in fact, increase the price of security. This is intuitive because, when the transmission lines are fully utilized, additional demand on either cheap or expensive sides does not change the cost of security. The reason is that when the transmission lines are fully utilized, the additional demand on either sides must be fulfilled locally. In this way, when more demand is distributed on the cheap node, the optimum economic dispatch cost is smaller than the case that the demand is distributed on the expensive node, hence the cost of security has a relatively bigger impact on the dispatch cost.

Finally, the following theorem is a direct consequence of the results in lemmas~\ref{lem:gen}, \ref{lem:dem2}, and~\ref{lem:dem}, and characterizes the worst-case price of security as defined in~\eqref{eq:pos_global}. 
\begin{theorem}
	\label{thm:pos}
	For the $2$-bus topology, $\textsf{PoS}(\ins)$ achieves its maximum value when $d_1 = 0$, $d_2 = f^{\textsf{ed}}$, and $\overline{q}_1 \geq f^{\textsf{ed}}$. Moreover, the maximum value is given by:
	\begin{equation}
	\label{eq:pos_opt2}
	\textsf{PoS} = \frac{\alpha_2}{\alpha_1} - \frac{(\alpha_2-\alpha_1) f^{\textsf{sc}}}{\alpha_1 f^{\textsf{ed}}}.
	\end{equation}
\end{theorem}

Theorem~\ref{thm:pos} states that the instance with the greatest price of security is such that all demand is at the expensive node and that demand is equal to the maximum flow from the cheap node to the expensive node in the economic dispatch problem. From the definitions in~\eqref{eq:fed} and~\eqref{eq:fsc}, it follows that $f^{\textsf{sc}} / f^{\textsf{ed}} \leq 1$. Observe that, as $f^{\textsf{sc}}/f^{\textsf{ed}} \uparrow 1$, the $\textsf{PoS}\downarrow 1$.

\begin{figure}[!t]
	\minipage{0.49\textwidth}
	\centering
	\includegraphics[scale=0.3]{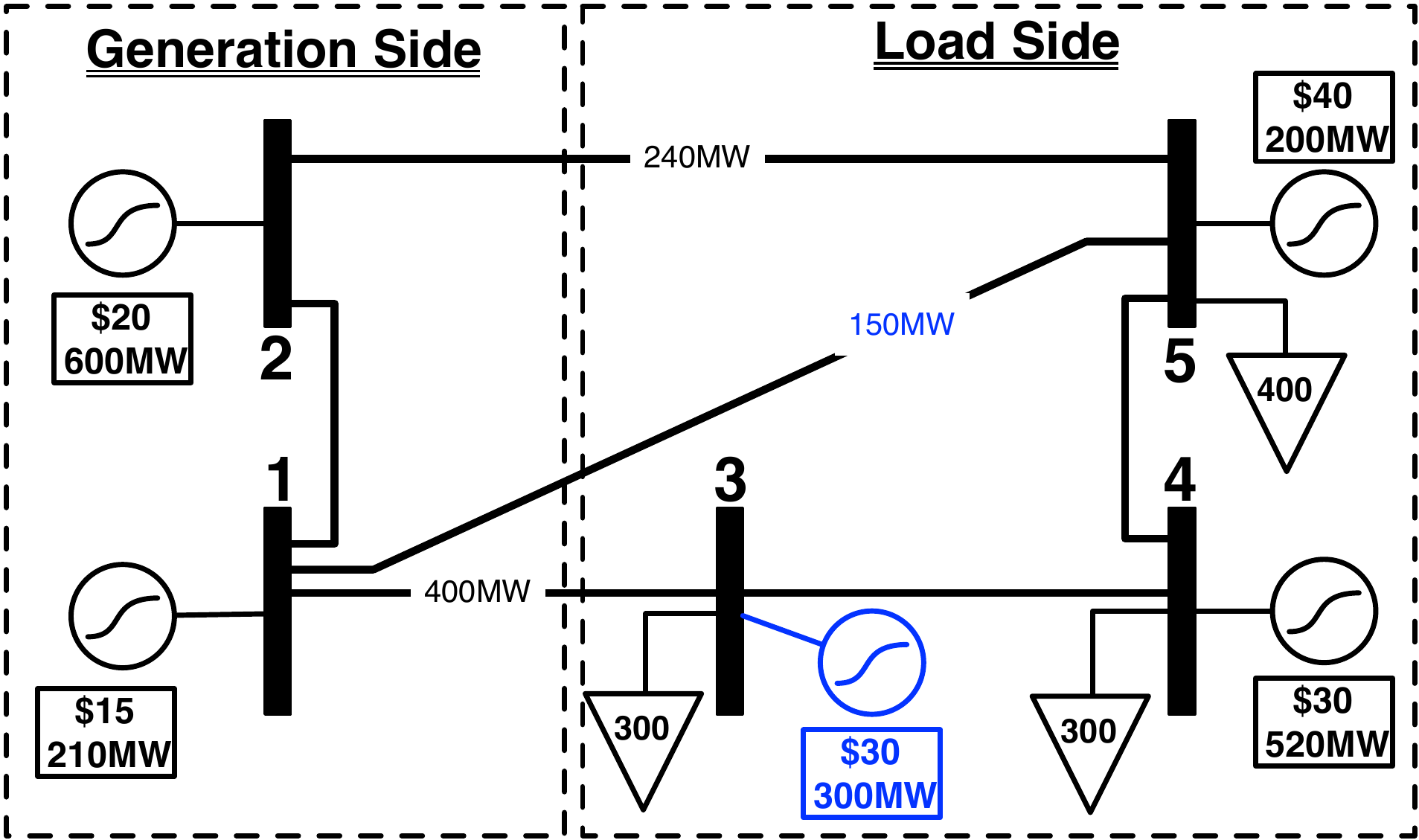}
	\caption{PJM $5$-bus topology, the modifications as compared to the original version are highlighted in blue.}
	\label{fig:5bus}
	\endminipage\hfill
\end{figure}
\begin{figure*}
	\centering
	\minipage{0.49\textwidth}
	\subfigure[$2$-bus topology]{
		\includegraphics[angle=0,scale=0.2]{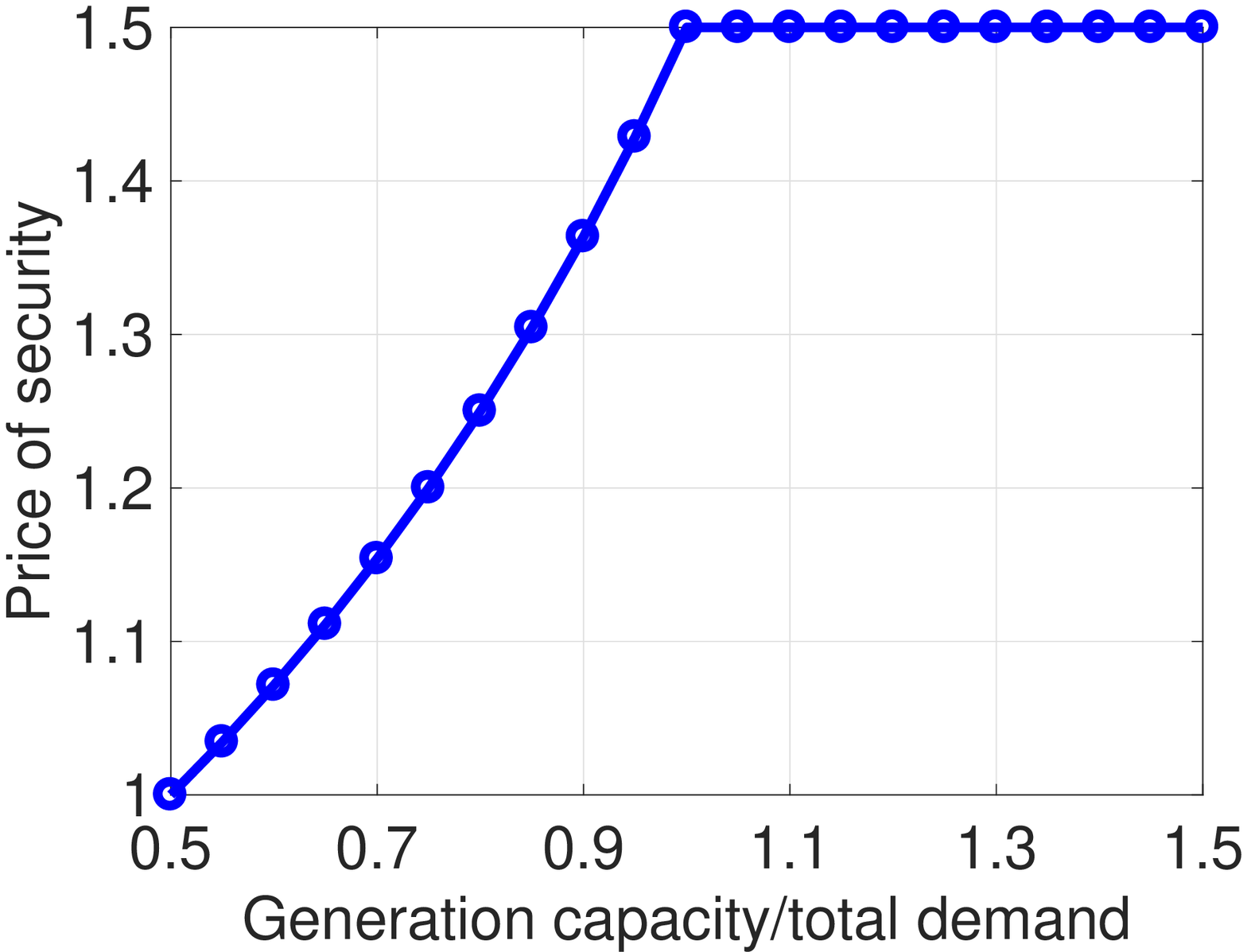}
		\vspace{-1\baselineskip}
		\label{fig:2bus_gen_cap_30steps}}
	\subfigure[PJM $5$-bus topology]{
		\vspace{0pt}\includegraphics[angle=0,scale=0.2]{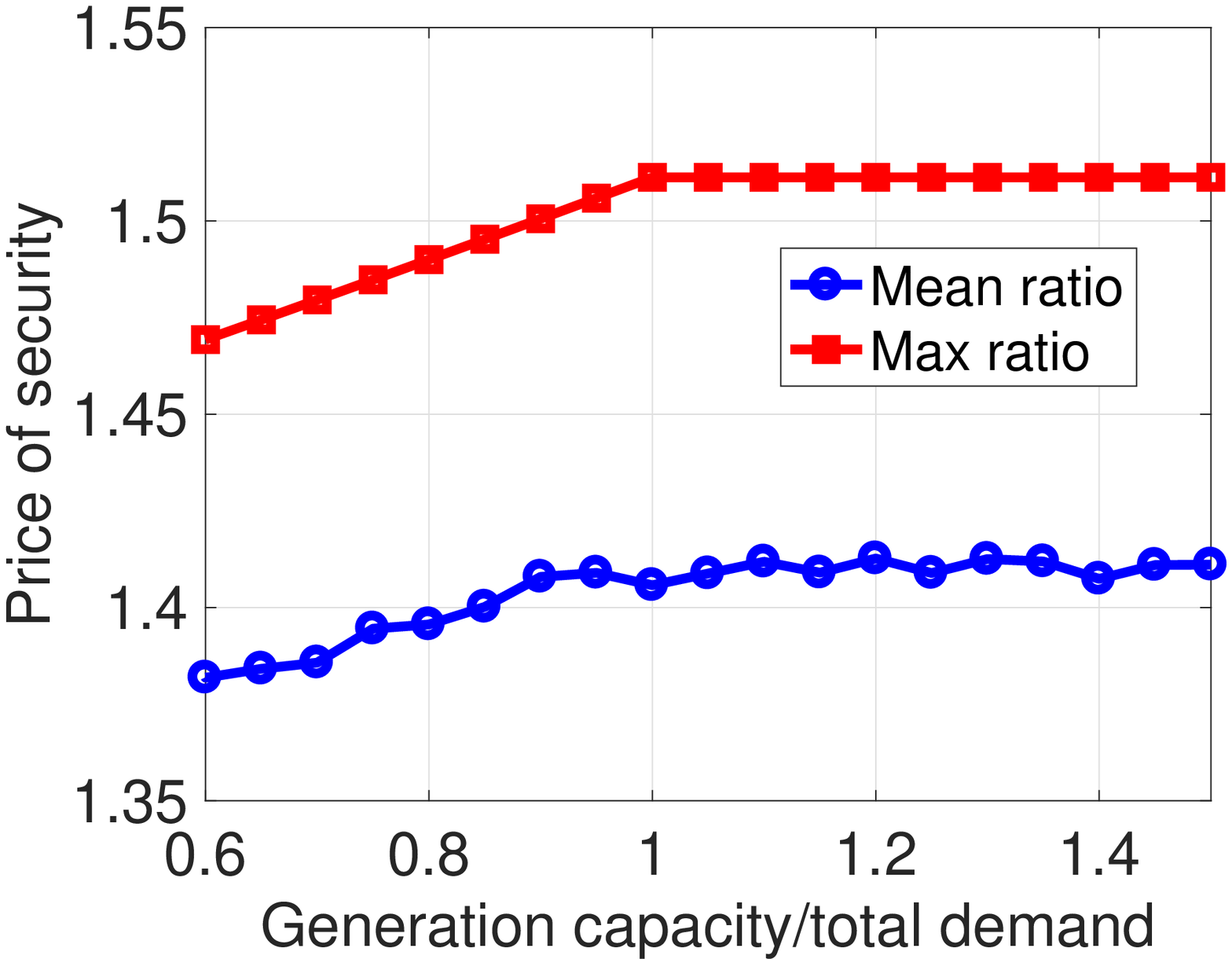}
		%\vspace{-1\baselineskip}5bus_gencap_500runs_20steps
		\label{fig:5bus_gencap_500runs_20steps}}
	%\vspace{-1\baselineskip}
	\caption{Price of security with fixed demand distribution and different generation capacities}
	\label{fig:lemma1}
	%\vspace{-1\baselineskip}
	\endminipage\hfill
	\centering
	\minipage{0.49\textwidth}
	\subfigure[$d_1 = [0,300)$, $d_2 = [0,300)$]{
		\includegraphics[angle=0,scale=0.2]{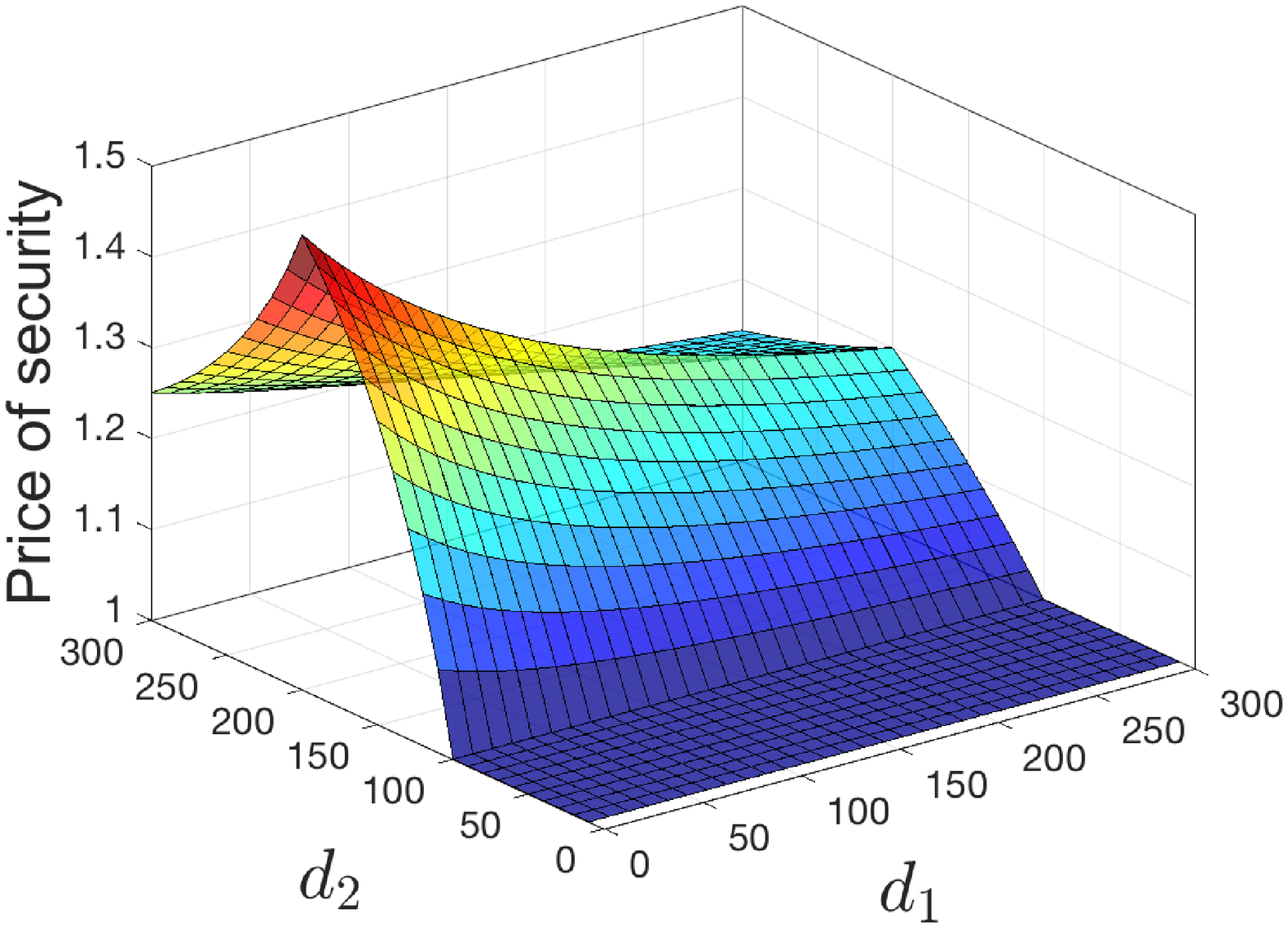}
		\label{fig:3d_2bus}}
	\subfigure[$d_1 = 0$, $d_2 = [0,300)$]{
		\includegraphics[angle=0,scale=0.2]{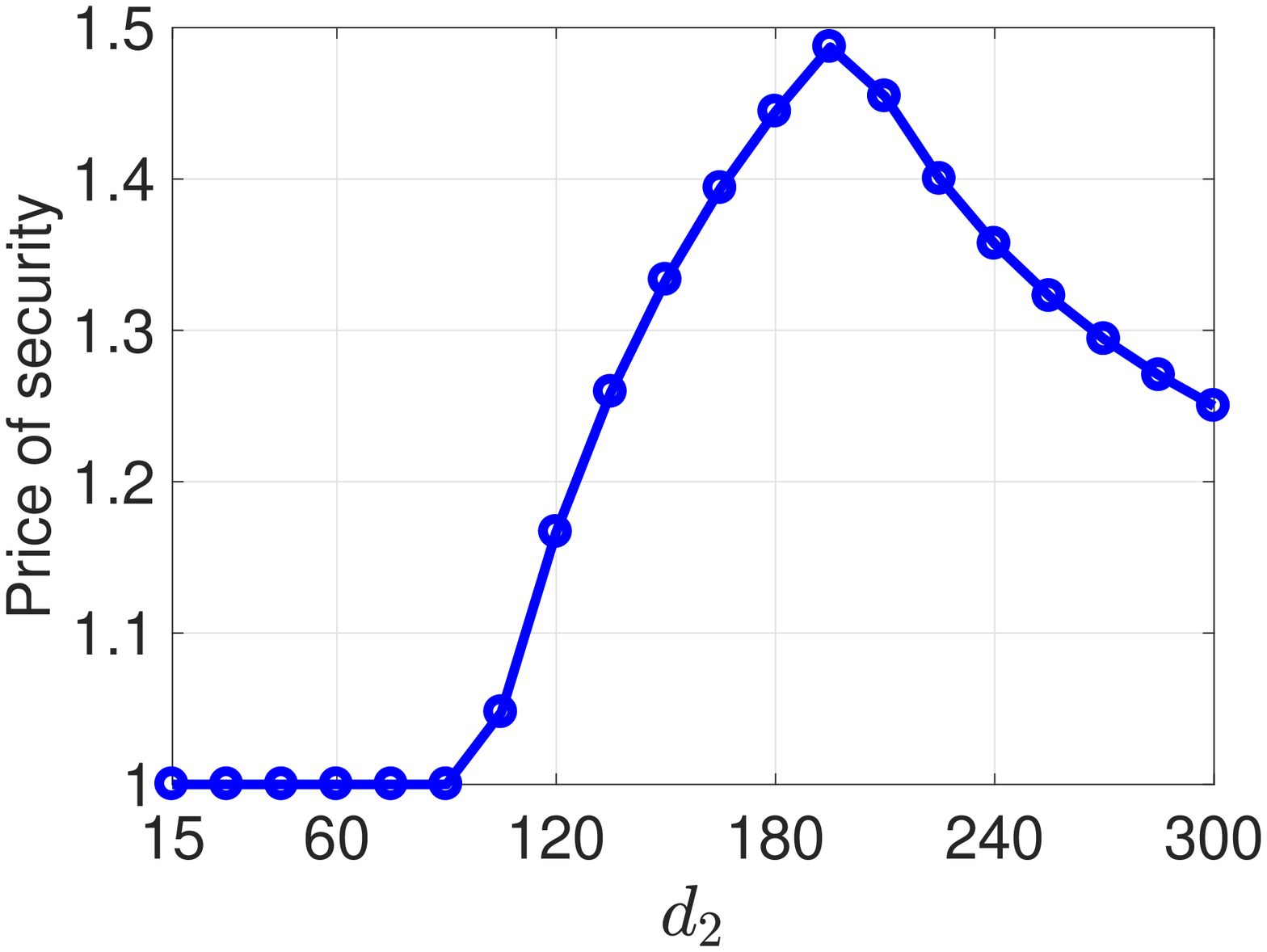}
		\label{fig:2bus_load_scaling_20steps_no_gen_cap}}
	\caption{Price of security for $2$-bus network in entire state space of demand distribution}
	\label{fig:2bus_thm2}
	\endminipage\hfill
	
\end{figure*}
\begin{figure*}[t]
	\centering
	\minipage{0.49\textwidth}
	\subfigure[$2$-bus topology]{
		\includegraphics[angle=0,scale=0.2]{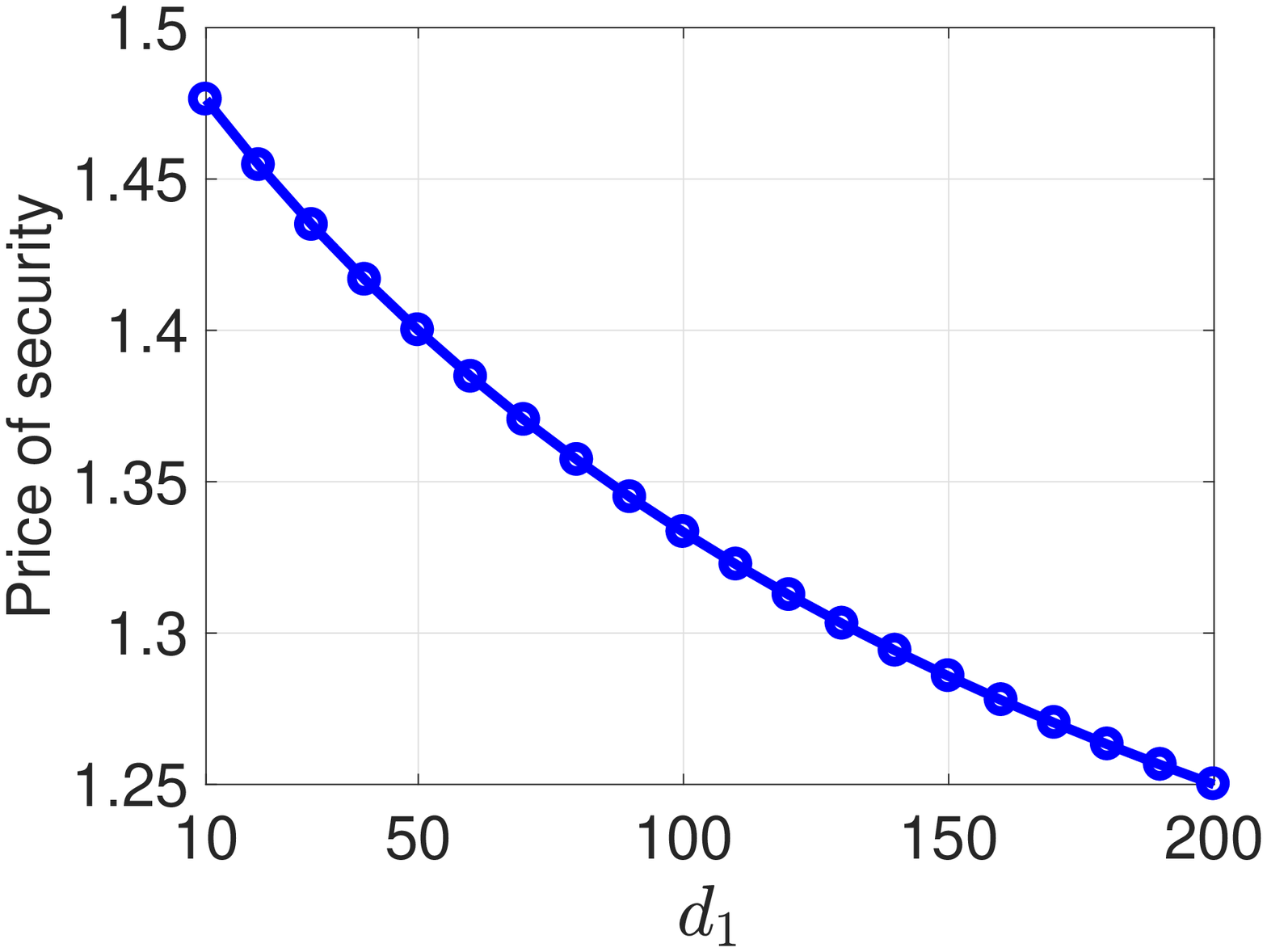}
		\vspace{-1\baselineskip}
		\label{fig:2bus_demand_20steps}}
	\subfigure[PJM $5$-bus topology]{
		\vspace{0pt}\includegraphics[angle=0,scale=0.2]{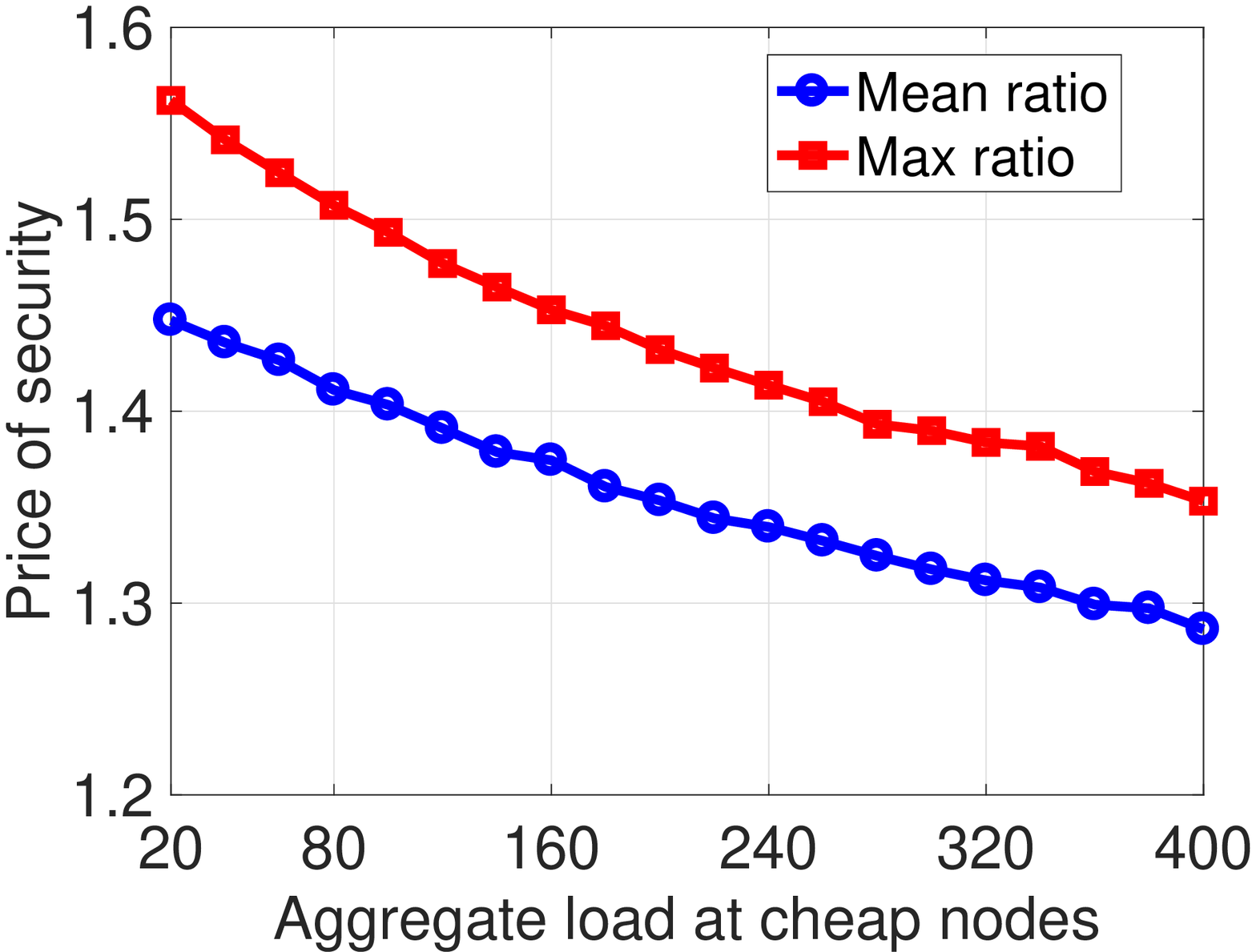}
		%\vspace{-1\baselineskip}
		\label{fig:5bus_load_scalingatcheap_500runs_20steps}}
	%\vspace{-1\baselineskip}
	\caption{Price of security with fixed expensive demand and different cheap demands}
	\label{fig:lemma2}
	%\vspace{-1\baselineskip}
	\endminipage\hfill
	\centering
	\minipage{0.49\textwidth}
	\subfigure[$2$-bus topology]{
		\includegraphics[angle=0,scale=0.2]{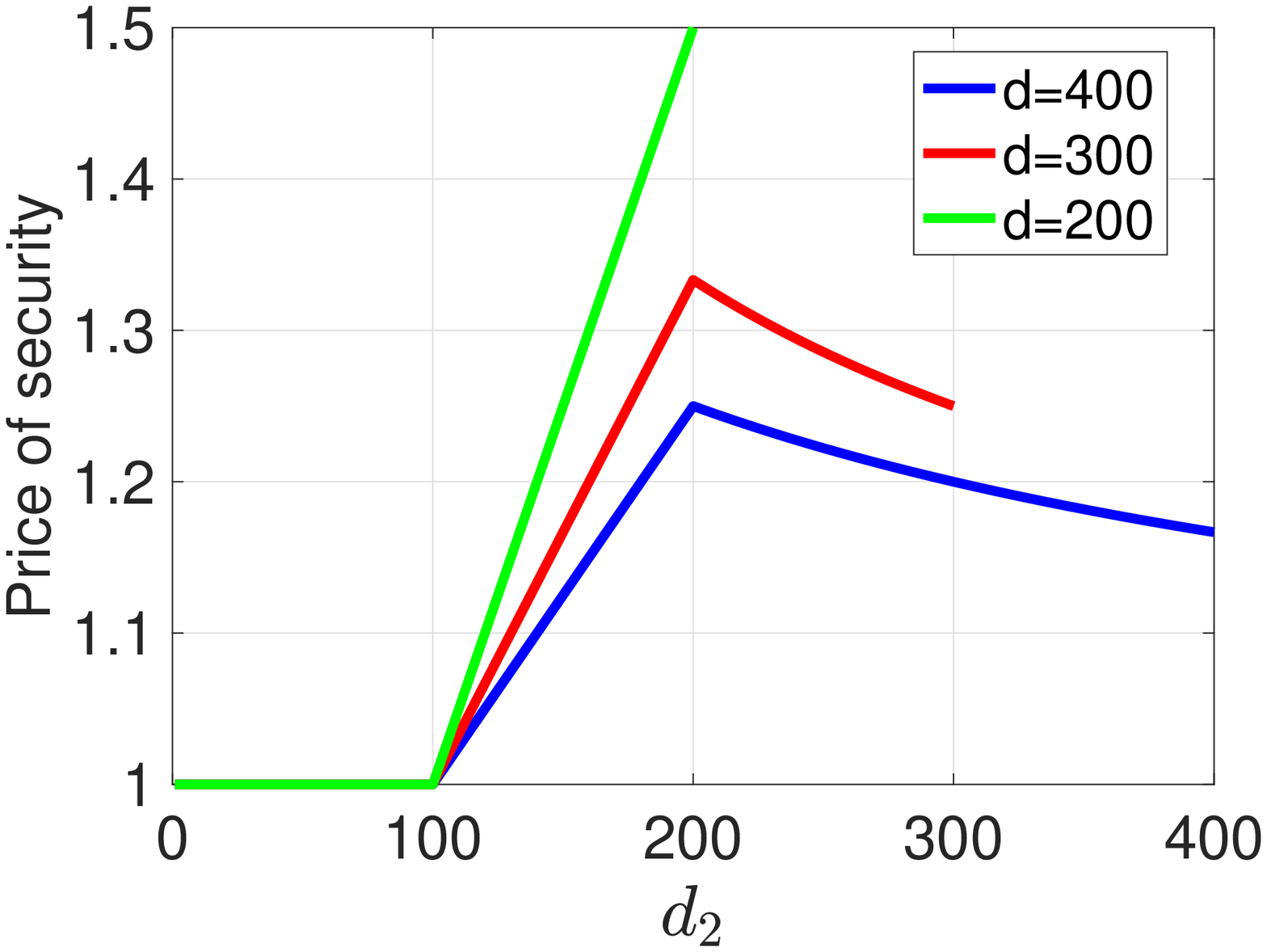}
		\vspace{-1\baselineskip}
		\label{fig:2bus_fixed_demand}}
	\subfigure[PJM $5$-bus topology]{
		\vspace{0pt}\includegraphics[angle=0,scale=0.2]{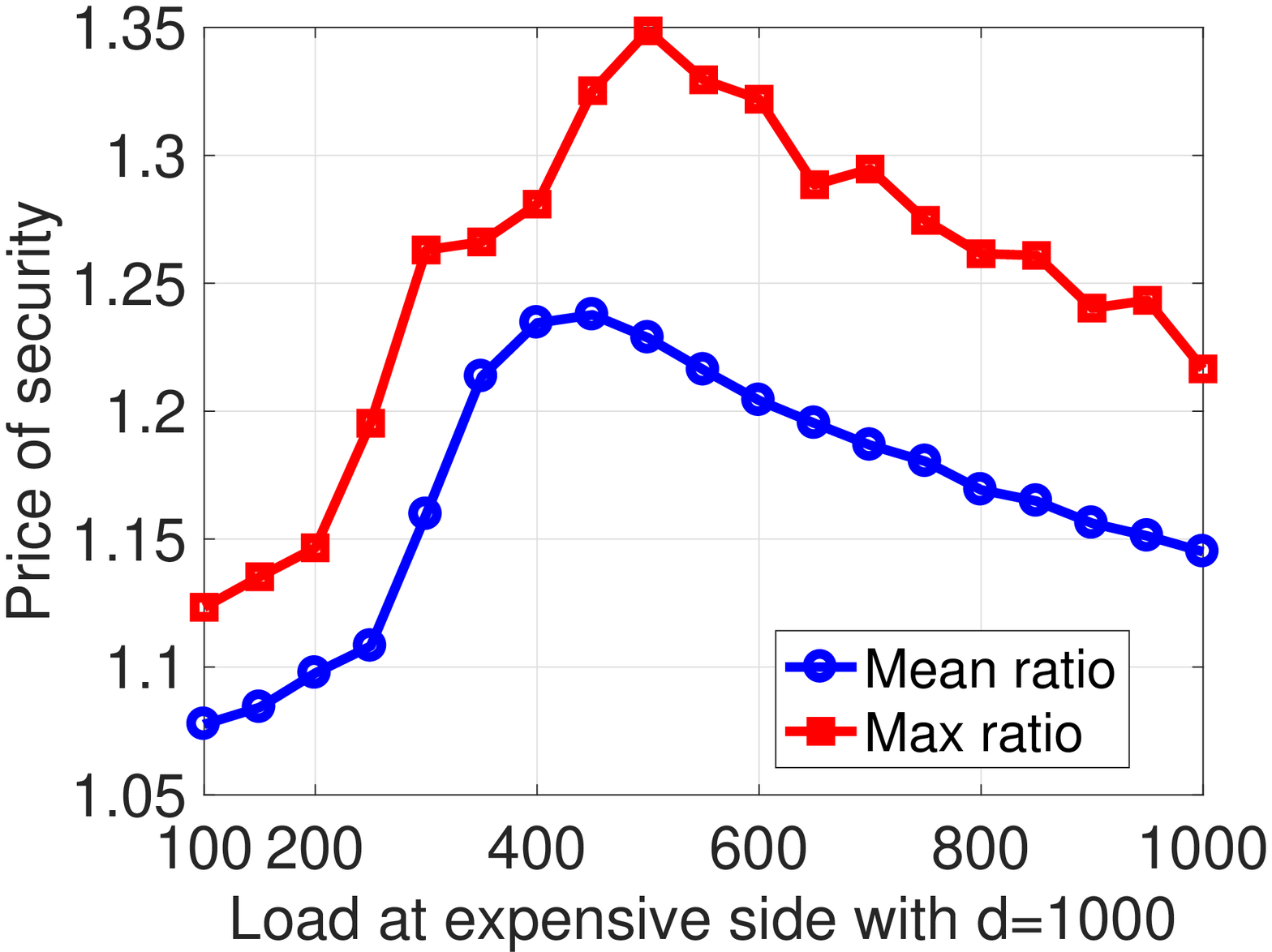}
		%\vspace{-1\baselineskip}
		\label{fig:5bus_load_scalingatcheap_500runs_20steps_no_gen_cap_min_max_v4}}
	%\vspace{-1\baselineskip}
	\caption{Price of security with fixed aggregate demand and different demand distribution}
	\label{fig:thm2}
	\vspace{-2.5mm}
	\endminipage\hfill
\end{figure*}
\section{Numerical Results}
In this section, we report the numerical results. We verify the analytical results for the $2$-bus topology. Also we investigate the validity of results for the PJM $5$-bus topology. 

\subsection{Experimental Setting and Testbeds}
We report our results for the simple $2$-bus topology~\cite{Monticelli1987Security} and the PJM $5$-bus topology~\cite{Li2010Small}. 
For $2$-bus case, we set ${\alpha_1 = 1, \alpha_2 = 2}$, ${\overline{f}_1 = \overline{f}_2 = 100, B_1 = B_2 = 1}$. In this way, we get ${f^{\textsf{ed}} = 200}$ and $f^{\textsf{sc}} = 100$. 
As depicted in Fig.~\ref{fig:5bus}, the PJM $5$-bus system is a tractable one in which the system is roughly divided into two regions: generation side and demand side. In generation side there are two generators at buses $1$ and $2$ with linear costs with $\alpha_1 = 15$ and  $\alpha_2=20$, which are generally cheaper than the generators at the demand side.\footnote{Note that in original PJM $5$-bus system there are two generators at bus $1$. In our model, we assume that at each bus there is one generator. Hence, we modify the system and consider one generator at bus $1$ with capacity equal to the aggregate capacity of the generators at bus $1$ in original version.}
We modify the test case in two ways: (i) the line limit of line $(1,5)$ is set to $150$MW; to make the test case more realistic in which each line comes with a maximum limit; and (ii) a generator is added at bus $3$; to ensure that the security-constrained problem is always feasible; by adding this generator the demand could be always fulfilled using local generations, regardless of line outage.
%To report statistical data for different demand profile, each data point in figures related to PJM $5$-bus system is average (or max) price of security of $500$ random runs, each of which with different demand profile with fixed overall characteristics.  

\begin{figure*}
	\centering
	\subfigure[With $150$MW link]{
		\vspace{0pt}\includegraphics[angle=0,scale=0.2]{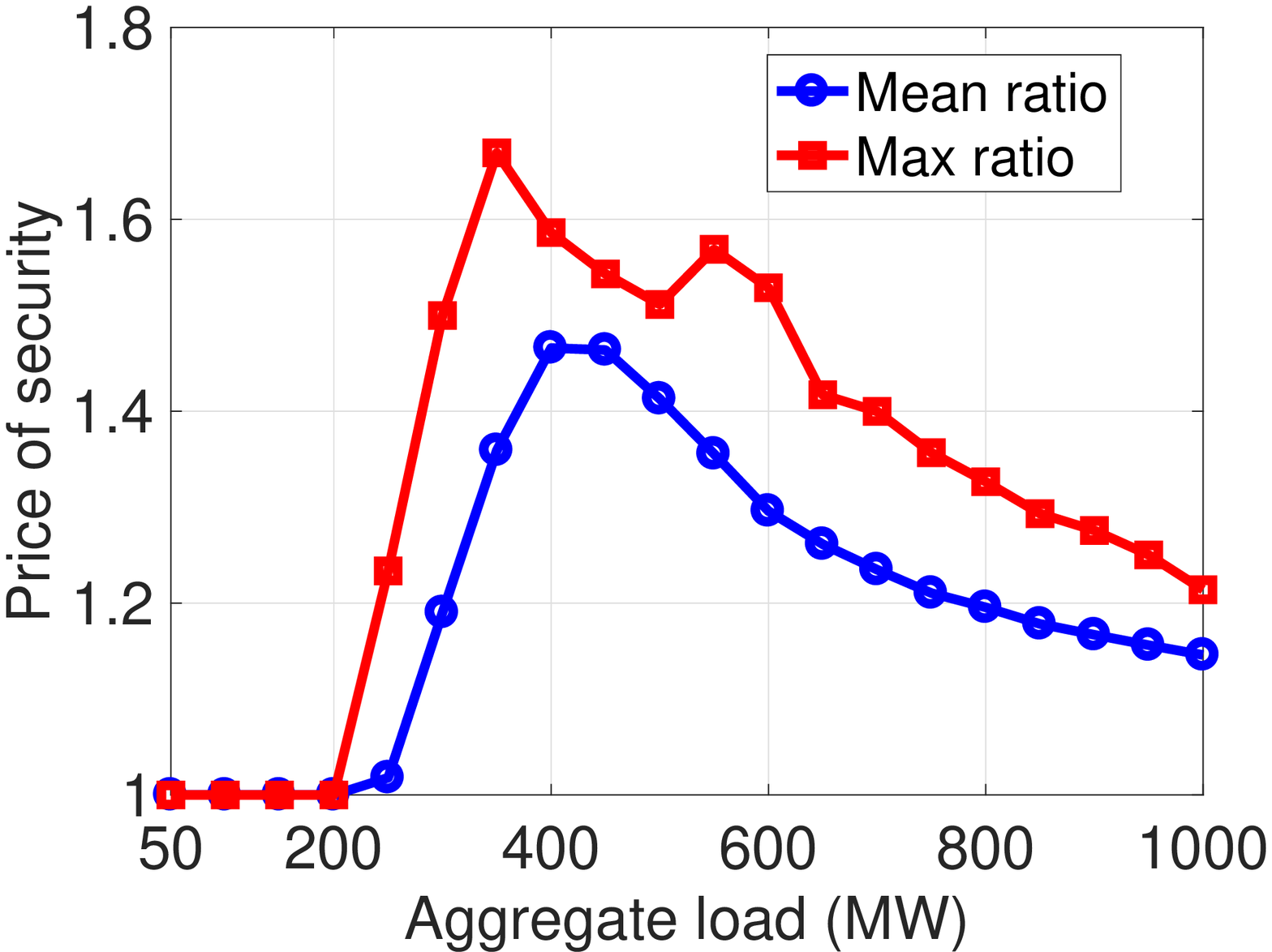}
		\label{fig:5bus_load_scaling_500runs_20steps_with150}}
	\subfigure[Without $150$MW link]{
		\vspace{0pt}\includegraphics[angle=0,scale=0.2]{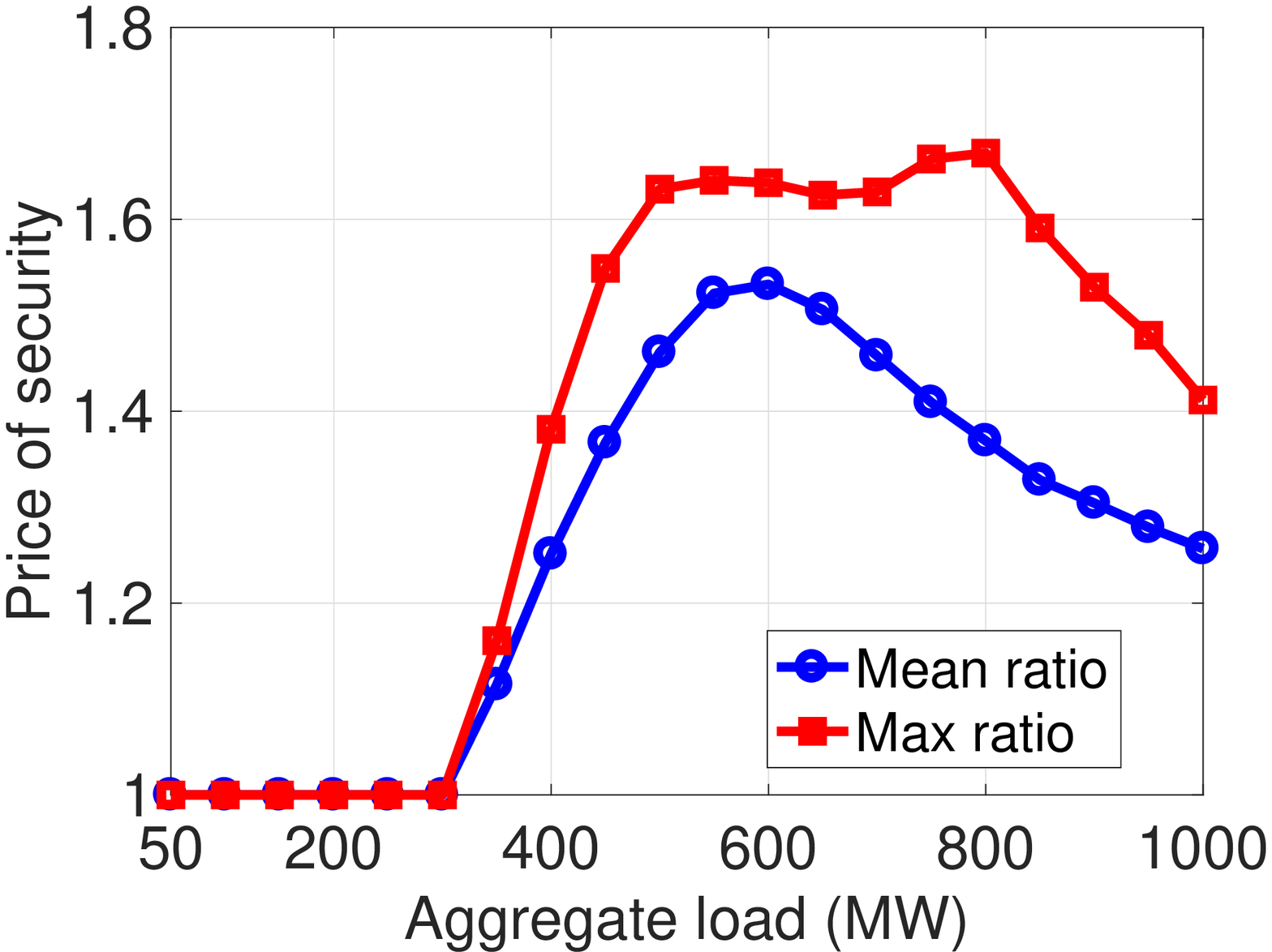}
		\label{fig:5bus_load_scaling_500runs_20steps_without150}}
	\subfigure[Without $150$MW link and normalized line limits]{
		\vspace{0pt}\includegraphics[angle=0,scale=0.2]{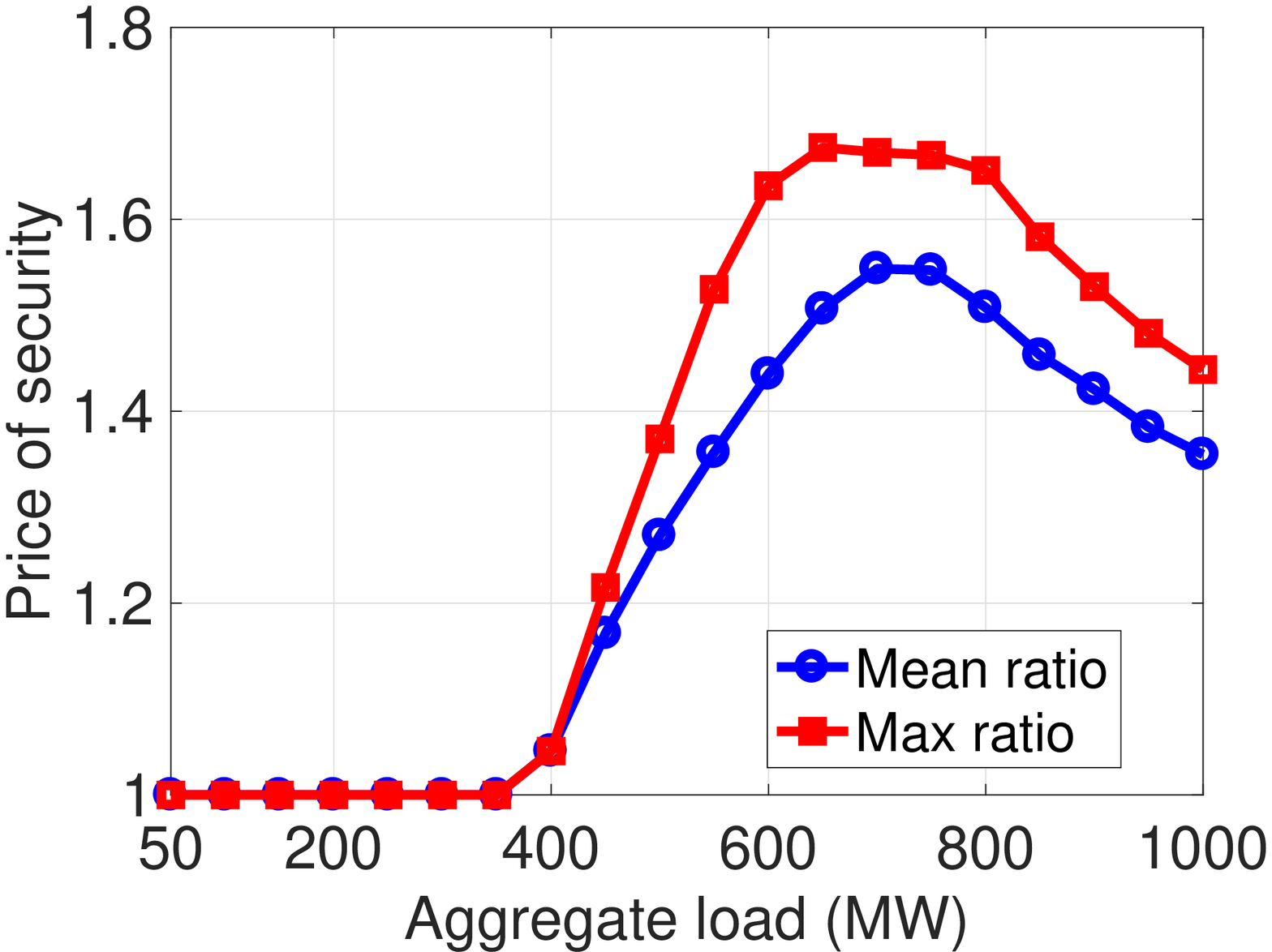}
		\label{fig:5bus_load_scaling_500runs_20steps_without150_normalized}}
	\subfigure[Without $150$MW link, normalized lines, and homogeneous costs]{
		\vspace{0pt}\includegraphics[angle=0,scale=0.2]{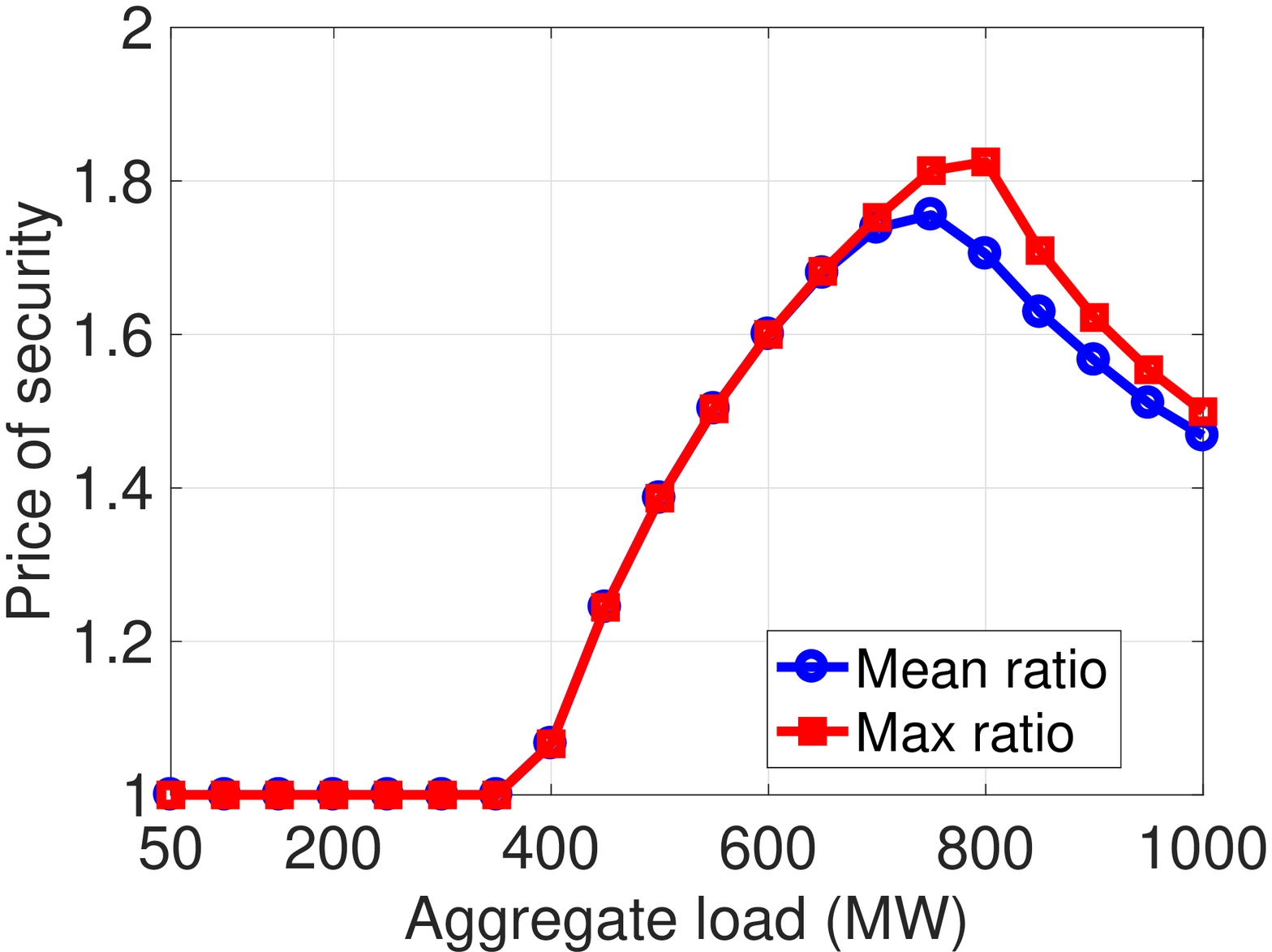}
		\label{fig:5bus_load_scaling_500runs_20steps_without150_normalized_homo}}	\caption{Results of PJM $5$-bus network with different values of demand}
	\label{fig:5bus_thm2}
\end{figure*}

\subsection{Impact of Generation Capacity}
In this experiment, we verify the claim in Lemma~\ref{lem:gen} using $2$-bus topology in Fig.~\ref{fig:2bus_gen_cap_30steps} and PJM $5$-bus in Fig.~\ref{fig:5bus_gencap_500runs_20steps}. 
In Fig.~\ref{fig:2bus_gen_cap_30steps}, while the demand is fixed ($d_1=0,d_2=200$), we change the generation capacity at generator $1$ from $50\%$ of demand ($\overline{q}_1=100$) to $150\%$ of the demand ($\overline{q}_1=300$). As stated in Lemma~\ref{lem:gen}, the price of security increases with the increase in generation capacity until the capacity reaches the total demand at $200$MW, and stays constant afterwards. 

In Fig.~\ref{fig:5bus_gencap_500runs_20steps}, the result of same experiment for PJM $5$-bus topology is reported. In this experiment, we change the generation capacity of cheap generators (at buses $1$ and $2$) in  $[0.6, 1.5]$ of total demand with step $0.05$. We note that even with fixed aggregate demand, the price of security changes with different demand distributions at nodes. Hence, we report the maximum and the average price of security of $500$ random runs each of which with different randomly generated demand profile on the expensive buses with fixed aggregate load. The result exhibits the same behavior as in $2$-bus topology and as the generation capacity increases the price of security increases. In summary, the results in Fig.~\ref{fig:lemma1} verify the analysis in Lemma~\ref{lem:gen}, which intuitively says that when the generation capacity is not the bottleneck, higher price of security is expected. Given the result in the this section, in the rest of the experiments, we relax the generation capacity of all generators in both topologies. 

\subsection{Impact of Demand Profile}
In this set of experiments, we verify the analytical results in lemmas~\ref{lem:dem2}, \ref{lem:dem} and Theorem~\ref{thm:pos}. 

\subsubsection{Price of Security in the Entire State Space} First, we focus on $2$-bus case and in Fig.~\ref{fig:2bus_thm2}, we report the price of security for the entire state space of demand distribution in cheap and expensive nodes. 
The most important observation is that the price of security is globally maximized when $d_1=0$ and $d_2=f^{\textsf{ed}} = 200$MW, which is consistent with the result in Theorem~\ref{thm:pos}.  Another observation in Fig.~\ref{fig:3d_2bus} is that given fixed demand at expensive side, the price of security achieves its maximum when $d_1=0$, which is consequence of Lemma~\ref{lem:dem2}. For better illustration, in Fig.~\ref{fig:2bus_load_scaling_20steps_no_gen_cap}, the price of security as a function of $d_2$ and for $d_1=0$ is reported. 

\subsubsection{Investigating the Result in Lemma~\ref{lem:dem2}}
In Fig.~\ref{fig:lemma2}, we investigate the result in Lemma~\ref{lem:dem2}, which says that given a fixed demand at the expensive node, the price of security decreases as the demand at the cheap node increases. For both topologies, we fix the (aggregate) demand at the expensive side and change the demand at cheap side. As shown in Fig.~\ref{fig:2bus_demand_20steps}, as the demand at cheap node $1$ increases, the price of security decreases. In Fig.~\ref{fig:5bus_load_scalingatcheap_500runs_20steps}, the aggregate demand in expensive nodes is fixed and equal to $400$MW and the aggregate load at two nodes $1$ and $2$ is changed from $20$ to $400$MW with step $20$ and at each point the average and maximum price of security of $500$ random runs are reported. The result clearly demonstrates that as the load at cheap nodes increases, the price of security decreases and when there is no demand at cheap nodes the maximum price of security is attained.

\subsubsection{Investigating the Result in Lemma~\ref{lem:dem}}
Next in Fig.~\ref{fig:thm2}, we investigate the statement in Lemma~\ref{lem:dem}. 
Toward this, we fix the aggregate demand and change the distribution of demand at cheap and expensive nodes. 
In Fig.~\ref{fig:2bus_fixed_demand}, we report the price of security, for $3$ different values of aggregate demand in $2$-bus topology. The result demonstrates that in all cases the price of security is maximized when the demand $d_2$ at the expensive node is equal to the maximum line capacity and the rest is at the cheap node, which is the result shown in Lemma~\ref{lem:dem}. In Fig.~\ref{fig:5bus_load_scalingatcheap_500runs_20steps_no_gen_cap_min_max_v4}, the result of the same experiment for PJM $5$-bus topology is reported. In this experiment, we fix the total demand at $d=1000$MW and change the distribution of load at two regions. The result shows the same general behavior with $2$-bus topology in the sense that the price of security reaches its maximum when roughly the demand at expensive node reaches the effective transmission capacity. 
However, different from explicit characterization of the maximum flow capacity in $2$-bus topology in~\eqref{eq:fed} and~\eqref{eq:fsc}, the effective transmission capacity in PJM $5$-bus topology is not straightforward to recognize. Furthermore, different peak values for the maximum and the average price of security imply that even with fixed aggregate demand at expensive side, price of security changes with different demand distribution. 
%
%\subsection{The Impact of Demand Profile in Worst-case Input}
%In this scenario our goal is to investigate the result in Theorem~\ref{thm:worst_pos} using $2$-bus topology in Fig.~\ref{fig:2bus_thm2} and PJM $5$-bus topology in Fig.~\ref{fig:2bus_thm2}.

\subsubsection{Insights for the Characterizing the Price of Security in General Networks}
Finally, we proceed to investigate the price of security for PJM $5$-bus topology in more details in the worst-case scenario similar to Fig.~\ref{fig:2bus_load_scaling_20steps_no_gen_cap} for $2$-bus topology, in which the demand at cheap side is zero. 
The first result for PJM $5$-bus topology is shown in~Fig.~\ref{fig:5bus_load_scaling_500runs_20steps_with150}. The notable observation is that the overall behavior is similar to the $2$-bus topology, since the price of security is $1$ at low load regimes when $d\leq 200$. Then, there is an increasing region (from $200$ to $\approx400$) in which the price of security increases with the increase of demand at expensive side, and eventually it achieves the maximum and then (when $d\geq400$) the price of security is decreasing. 

The result, however, is different from $2$-bus topology in a way that the \textit{critical points} (the point at which the price begins to increase, and the one at which the price takes it maximum) are not straightforward function of line properties. Recall that these points are characterized explicitly in~\eqref{eq:fed} and~\eqref{eq:fsc} for $2$-bus topology. 
%In particular in~\eqref{eq:fed}, when ${\overline{f}_1/B_1=\overline{f}_2/B_2}$, we get ${f^{\textsf{ed}} = \overline{f}_1+\overline{f}_2}$, and in worst case, i.e., $d=f^{\textsf{ed}}$. 
Fig.~\ref{fig:5bus_load_scaling_500runs_20steps_with150} shows that in worst-case, the aggregate demand is less than aggregate line limit from the cheap side to expensive side that is $790$MW. Thus, this result shows that characterizing the worst case demand profile is more challenging in PJM $5$-bus topology. 

To investigate how network topology and line characteristics can impact the two aforementioned critical points of the price of security, in three consecutive steps, we simplify PJM $5$-bus topology to be similar to $2$-bus case. Toward this, we first remove the line $(1,5)$ with capacity $150$MW (reported in Fig.~~\ref{fig:5bus_load_scaling_500runs_20steps_without150}), second, we normalized the link capacities such that ${f_{(1,3)}/B_{(1,3)}=f_{(2,5)}/B_{(2,5)}}$ (reported in Fig.~\ref{fig:5bus_load_scaling_500runs_20steps_without150_normalized}); recall that $f^{\textsf{ed}}$ in $2$-bus topology is maximized when ${\overline{f}_1/B_1=\overline{f}_2/B_2}$; third, we set the homogeneous generator costs at $\$15$ at generation side and $\$40$ at demand side (reported in Fig.~\ref{fig:5bus_load_scaling_500runs_20steps_without150_normalized_homo}). 

The observations are as follows: (i) price of security at the second critical point where the price of security reaches its maximum, increases as the network topology simplifies ($1.47 \rightarrow 1.53 \rightarrow 1.55 \rightarrow 1.75$); (ii) the aggregate demand at which the first critical point occurs, i.e, the point where the price of security starts to increase, increases as the networks simplifies ($200\text{MW} \rightarrow 300 \rightarrow 350 \rightarrow 350$); and finally, 
the aggregate demand at which the the price of security maximizes also increases ($400\text{MW} \rightarrow 600 \rightarrow 700 \rightarrow 750$). These observations demonstrate that the worst-case aggregate demand that leads to maximum price of security depends on several characteristics of topology, transmission lines, and cost functions.

\section{Conclusions and Future Directions}
In this paper, we tackle a novel and important problem on understanding the economic cost of incorporating security constraints in economic dispatch. We introduce the notion of price of security as a metric that formally characterizes the economic inefficiency of ensuring $N-1$ security. Focusing on security to line outages in a $2$-bus-$2$-line topology, we investigate the impact of generation availability and demand distribution on the price of security. In addition, we show that the price of security is greatest when the entire demand is equal to the maximum flow of the lines and is placed at the expensive side of the network. Extensive experimental results on the PJM 5-bus system show that some of our theoretical observations manifest in more general settings. 

%This paper studies the properties that can impact the price of security, i.e., the economic loss due to respect security constraints in economic dispatch problem. 
%By analysis on simple 2-bus system and considering the entire state space for input instances, it demonstrates that in the cases that the capacity of cheap generator is larger than the demand and the entire demand is at the bus with expensive generator lead to worst price of security. 
%Then, the result is extended to more general case, which is a common case in practice, and the conditions are derived that ...

As future work, we plan to extend our theoretical results to more general networks. The starting point, perhaps, is to consider a network that can be divided into two regions, one cheap and one expensive. Our theoretical results indicate that the price of security may depend critically on the maximum flow from the cheap to the expensive region. Therefore, the challenge is to characterize the maximum flow. Our experiments demonstrate that the price of security also depends significantly on the topology within the regions, demand distributions, and line characteristics. Another direction is to study the price of security for corrective security-constrained economic dispatch and contrast the latter with preventive approaches to understand the tradeoffs between operating costs and security benefits.

%At the cheap side, there are several homogeneous generators in a set of buses with arbitrary topology. Then, through a set of transmission lines with limited aggregate thermal limits, the power is transmitted the expensive side, where the total demand is distributed arbitrarily in multiple nodes. Then, the goal is to characterize the effective transmission capacity of the lines in the middle. Our experiments demonstrate that it highly depends on the topology, demand distribution, and line characteristics. Another direction is to study the price of security for the corrective approach in security-constrained economic dispatch. In this way, the superiority of corrective approaches as compared to preventive approaches could be characterized. Overall, we believe that characterizing the price of security in different approaches is an important problem and call for further investigation. 

%In addition, in practice, the preventive formulation comes with an overload degree in which in case of line outage the lines could be overloaded for a while. Another future direction is to characterize the price of security in overloaded preventive security-constrained economic dispatch. 

\bibliographystyle{ieeetr}

\appendix

\subsection{Proof of Lemma~\ref{lem:gen}}
We consider two cases. 

\textbf{Case 1:} Constraint~\eqref{eq:gen_cap_2bus} is not active in the optimal solution  of instance $\ins'$ in \textsf{ED-2b}, i.e., $q_1^{'\star} < \overline{q}'_1$, where $q_1^{'\star}$ is the optimal output of generator $1$ in \textsf{ED-2b}. This means that even though the capacity of generator $1$ is limited in instance $\ins'$, i.e., $\overline{q}'_1 \leq d$, the other constraints in~\eqref{eq:sup_dem_bal_2bus},~\eqref{eq:line_cap_2bus}, and~\eqref{eq:line_cap_2bus_2} hinder the generation of generator $1$ at full capacity. Consequently, the generation capacity of cheap generator is not a bottleneck in problem, thereby increasing it in instance $\ins$ would not change the optimal value and solution of \textsf{ED-2b}, i.e., $c^{\star}_{\textsf{ed}}(\ins) = c^{\star}_{\textsf{ed}}(\ins')$. Similarly, we have ${c^{\star}_{\textsf{sc}}(\ins) = c^{\star}_{\textsf{sc}}(\ins')}$, since the constraint set of \textsf{SCED-2b} is more restricted than the set of \textsf{ED-2b}, and if~\eqref{eq:gen_cap_2bus} is not active in \textsf{ED-2b}, the equivalent constraint~\eqref{eq:gen_cap_2bus_sc} in \textsf{SCED-2b} is not active as well. Putting together, we have $\textsf{PoS}(\ins) = \textsf{PoS}(\ins')$. 

\textbf{Case 2:} Constraint~\eqref{eq:gen_cap_2bus} is active in the optimal solution  of instance $\ins'$ in the \textsf{ED-2b}, i.e., $q_1^{'\star} = \overline{q}'_1$. In this case, we prove $\textsf{PoS}(\ins') \leq \textsf{PoS}(\ins)$.  

To prove, we write the optimal cost of instance $\ins'$ as a function of optimal cost of instance $\ins$ in both original and security constrained problems. We have the following costs for instance $\ins$:
\begin{eqnarray}
c^{\star}_{\textsf{ed}}(\ins) &=& \alpha_1 q_1^{\star} + \alpha_2 (d-q_1^{\star}), \label{eq:ed_cost_sigma}\\
c^{\star}_{\textsf{sc}}(\ins) &=& \alpha_1 (q_1^{\star} - a) + \alpha_2 (d-q_1^{\star}+a), \label{eq:sc_cost_sigma}
\end{eqnarray}
where $q_1^{\star}$ is the optimal output of generator $1$ and $d=d_1+d_2$ and $a \geq 0$ is the amount of reduction in generation output of generator $1$ due to respecting the security constraints in instance $\ins$.

On the other hand in instance $\ins'$, we have $q_1^{'\star} = \overline{q}'_1$, so we get the following cost $c^{\star}_{\textsf{ed}}(\ins)$:
\begin{eqnarray}
c^{\star}_{\textsf{ed}}(\ins') &=& \alpha_1 \overline{q}'_1 + \alpha_2 (d - \overline{q}'_1) \nonumber \\
&=& \alpha_1 (q_1^{\star}-(q_1^{\star}-\overline{q}'_1)) + \alpha_2 (d-q_1^{\star} + q_1^{\star} - \overline{q}'_1) \nonumber \\
&=& \alpha_1 q_1^{\star} + \alpha_2 (d-q_1^{\star}) + (\alpha_2 - \alpha_1) (q_1^{\star} - \overline{q}'_1)\nonumber \\
&=& c^{\star}_{\textsf{ed}}(\ins) + A, 
\end{eqnarray}
where $A = (\alpha_2 - \alpha_1) (q_1^{\star} - \overline{q}'_1)$. Note that $A \geq 0$ since $\alpha_2 \geq \alpha_1$ and $q_1^{\star}\geq \overline{q}'_1$.
Similarly, we have 
\begin{eqnarray}
c^{\star}_{\textsf{sc}}(\ins') &=& \alpha_1 (\overline{q}'_1 -b) + \alpha_2 (d - \overline{q}'_1 + b) \nonumber \\
&=& \alpha_1 (q_1^{\star}-(q_1^{\star}-\overline{q}'_1) - b) \nonumber\\
&&+ \alpha_2 (d - q_1^{\star}+q_1^{\star} - \overline{q}'_1 + b) \nonumber \\
&=& \alpha_1 (q_1^{\star}-a) + \alpha_2 (d-q_1^{\star}+a) \nonumber\\
&&+ (\alpha_2 - \alpha_1) (q_1^{\star} - \overline{q}'_1+b-a)\nonumber \\
&=& c^{\star}_{\textsf{sc}}(\ins) + A + B, 
\end{eqnarray}
where $b \geq 0$ is the amount of reduction in generation output of generator $1$ due to respecting security constraints in instance $\ins'$ and $B = (\alpha_2 - \alpha_1) (b-a)$. We note that $B\leq 0$ since $\alpha_2 \geq 0$ and $b \leq a$. Note that $b$ could be at most $a$ which is the amount of reduction in output of generator $1$ to respect security constraints. 
Now we proceed to calculate the price of security for input instance $\ins'$ as follows
\begin{eqnarray*}
	\textsf{PoS}(\ins') &=& \frac{c^{\star}_{\textsf{sc}}(\ins')}{c^{\star}_{\textsf{ed}}(\ins')}\\ &=& \frac{c^{\star}_{\textsf{sc}}(\ins) + A + B}{c^{\star}_{\textsf{ed}}(\ins) + A} \\
	&\leq& \frac{c^{\star}_{\textsf{sc}}(\ins)}{c^{\star}_{\textsf{ed}}(\ins)} = \textsf{PoS}(\ins),
\end{eqnarray*}
where the inequality follows from $\frac{x}{y} \geq \frac{x+z}{y+z}$, if $x\geq y$ and $z\geq 0$.  Putting together both cases the result is proven. 

\subsection{Proof of Lemma~\ref{lem:dem2}}
We claim that $c^{\star}_{\textsf{ed}}(\ins') = c^{\star}_{\textsf{ed}}(\ins) + \alpha_1 (d'_1 - d_1)$. The reason is that the increase in demand, i.e., $(d'_1 - d_1)$, is in the side of cheap node, so it can be locally satisfied by the cheap generator at node $1$ with the minimum cost of $\alpha_1 (d'_1 - d_1)$. Similarly, we get ${c^{\star}_{\textsf{sc}}(\ins') = c^{\star}_{\textsf{sc}}(\ins) + \alpha_1 (d'_1 - d_1)}$, since the additional security constraints do not impact on the ability of local generation. Now, we have 
\begin{eqnarray*}
	\textsf{PoS}(\ins') &=& \frac{c^{\star}_{\textsf{sc}}(\ins')}{c^{\star}_{\textsf{ed}}(\ins')}= \frac{c^{\star}_{\textsf{sc}}(\ins) + \alpha_1 (d'_1 - d_1)}{c^{\star}_{\textsf{ed}}(\ins) + \alpha_1 (d'_1 - d_1)} \\
	&\leq& \frac{c^{\star}_{\textsf{sc}}(\ins)}{c^{\star}_{\textsf{ed}}(\ins)} = \textsf{PoS}(\ins),
\end{eqnarray*}
where the last steps follows again from $\frac{x}{y}\geq\frac{x+z}{y+z}$ when $x\geq y$ and $z\geq 0$.
\subsection{Proof of Lemma~\ref{lem:dem}}
Given fixed $d=d_1+d_2$, our goal is to find the maximum value of~\eqref{eq:pos2bus}.
Since the equation of price of security is piecewise smooth, we proceed to find the maximum values in each piece separately. 

\textbf{Case 1:}  $d \leq f^\textsf{sc} \leq f^\textsf{ed}$: in this case regardless of how the demand is distributed at nodes $1$ and $2$, we have ${c^{\star}_{\textsf{ed}}(\ins) = c^{\star}_{\textsf{sc}}(\ins) = \alpha_1 d}$, hence $\textsf{PoS}(\ins) = 1$. 

\textbf{Case 2:} $f^\textsf{sc} \leq d \leq f^\textsf{ed}$: in this case we have ${c^{\star}_{\textsf{ed}}(\ins) = \alpha_1 d}$, 
and ${c^{\star}_{\textsf{sc}}(\ins) =\alpha_1  (d-d_2+f^{\textsf{sc}}) + \alpha_2  (d_2-f^{\textsf{sc}})}$
hence 
$$\textsf{PoS}(\ins) = \frac{\alpha_1  (d-d_2+f^{\textsf{sc}}) + \alpha_2  (d_2-f^{\textsf{sc}})}{\alpha_1 d}.$$
By taking the derivative we get
$$\frac{\partial \textsf{PoS}(\ins)}{\partial d_2} = \alpha_2-\alpha_1 > 0,$$ hence  $\textsf{PoS}(\ins)$ is strictly increasing function and takes its maximum when $d_2=d=\min\{d,f^{\textsf{ed}}\}$. 

\textbf{Case 3:} $f^\textsf{sc} \leq f^\textsf{ed} \leq d$: we claim that ${\ins = \left(\mathbf{\overline{q}} =\{\overline{q}_1\geq d, \overline{q}_2\}, \mathbf{d} = (d - f^{\textsf{ed}}, f^{\textsf{ed}}\})\right)}$ maximizes the price of security. In this case we have 
$$c^{\star}_{\textsf{ed}}(\ins) = \alpha_1 d,$$
and 
$${c^{\star}_{\textsf{sc}}(\ins) = \alpha_1 (d-f^{\textsf{ed}}+f^{\textsf{sc}}) + \alpha_2 (f^{\textsf{ed}}-f^{\textsf{sc}})}.$$
Now, we prove the claim by contradiction. Assume that there is an instance $\ins'$ where $\textsf{PoS}(\ins') > \textsf{PoS}(\ins)$. Given the state space, $\ins'$ must be in one of the following forms: 
\begin{eqnarray}
\label{eq:ins1}
\ins' = \left(\mathbf{\overline{q}} =\{\overline{q}_1\geq d, \overline{q}_2\}, \mathbf{d} = (d - f^{\textsf{ed}}+\epsilon, f^{\textsf{ed}}-\epsilon\})\right)\\
\label{eq:ins2}\ins' = \left(\mathbf{\overline{q}} =\{\overline{q}_1\geq d, \overline{q}_2\}, \mathbf{d} = (d - f^{\textsf{ed}}-\epsilon, f^{\textsf{ed}}+\epsilon\})\right).
\end{eqnarray}
We first compare the price of security of instance in~\eqref{eq:ins1}. We have 
$$c^{\star}_{\textsf{ed}}(\ins') = \alpha_1 d = c^{\star}_{\textsf{ed}}(\ins),$$
and 
\begin{eqnarray*}
	c^{\star}_{\textsf{sc}}(\ins') &=& \alpha_1 (d-f^{\textsf{ed}}+\epsilon+f^{\textsf{sc}}) + \alpha_2 (f^{\textsf{ed}}-\epsilon-f^{\textsf{sc}})\\
	&=& c^{\star}_{\textsf{sc}}(\ins) - (\alpha_2-\alpha_1)\epsilon < c^{\star}_{\textsf{sc}}(\ins)
\end{eqnarray*}
Apparently, we have $\textsf{PoS}(\ins') \!<\! \textsf{PoS}(\ins)$, which is contradiction. 
Similarly, we compare the price of security of instance in~\eqref{eq:ins2}, with the original instance $\ins$. We have 
\begin{eqnarray*}
	c^{\star}_{\textsf{ed}}(\ins') &=& \alpha_1 (d-\epsilon) + \alpha_2 \epsilon\\
	&=& c^{\star}_{\textsf{ed}}(\ins) + (\alpha_2-\alpha_1)\epsilon > c^{\star}_{\textsf{sc}}(\ins)
\end{eqnarray*}
and 
\begin{eqnarray*}
	c^{\star}_{\textsf{sc}}(\ins') &=& \alpha_1 (d-f^{\textsf{ed}}-\epsilon+f^{\textsf{sc}}) + \alpha_2 (f^{\textsf{ed}}+\epsilon-f^{\textsf{sc}})\\
	&=& c^{\star}_{\textsf{sc}}(\ins) + (\alpha_2-\alpha_1)\epsilon > c^{\star}_{\textsf{sc}}(\ins)
\end{eqnarray*}
Hence for the price of security we get 
\begin{eqnarray*}
	\textsf{PoS}(\ins') &=& \frac{c^{\star}_{\textsf{sc}}(\ins')}{c^{\star}_{\textsf{ed}}(\ins')}\\ &=& \frac{c^{\star}_{\textsf{sc}}(\ins) + (\alpha_2-\alpha_1)\epsilon}{c^{\star}_{\textsf{ed}}(\ins) + (\alpha_2-\alpha_1)\epsilon} \\
	&<& \frac{c^{\star}_{\textsf{sc}}(\ins)}{c^{\star}_{\textsf{ed}}(\ins)} = \textsf{PoS}(\ins),
\end{eqnarray*}
which is again a contrary. Hence the proof is completed.

\end{document}